\documentclass[11pt]{article}
\usepackage{amssymb}
\def\lddots{\mathinner{\mkern1mu\raise1pt\hbox{.}\mkern2mu
\raise4pt\hbox{.}\mkern2mu\raise7pt\vbox{\kern7pt\hbox{.}}\mkern1mu}}
\makeatletter
\def\numberbysection{\@addtoreset{equation}{section}
\def\theequation{\thesection.\arabic{equation}}}
\makeatother

\usepackage{epic}
\usepackage{eepic}
\usepackage{graphicx}
\usepackage{latexsym}
\usepackage{amssymb}
\numberbysection

\newcommand{\be}{\begin{eqnarray}}
\newcommand{\ee}{\end{eqnarray}}
\newcommand{\non}{\nonumber}

\newcommand{\tr}{\mathop{\rm tr}\nolimits}

\begin{document}

\begin{titlepage}

\vskip 0.4cm

\strut\hfill

\vskip 0.8cm

\begin{center}


{\bf {\LARGE The open XXZ and associated models at $q$ root of unity}}

\vspace{10mm}

{\Large Anastasia Doikou\footnote{e-mail: doikou@bo.infn.it}}

\vspace{14mm}

\it{INFN Section of Bologna, Physics Department, University of Bologna\\
Via Irnerio 46, 40126 Bologna, Italy}

\end{center}

\vfill

\begin{abstract}

The generalized open XXZ model at $q$ root of unity is considered.
We review how associated models, such as the $q$ harmonic
oscillator, and the lattice sine-Gordon and Liouville models are
obtained. Explicit expressions of the local
Hamiltonian of the spin ${1 \over 2}$ XXZ spin chain coupled to
dynamical degrees of freedom at the one end of the chain are provided.
Furthermore, the boundary non-local charges are given for the
lattice sine Gordon model and the $q$ harmonic oscillator with
open boundaries. We then identify the spectrum and the
corresponding Bethe states, of the XXZ and the q harmonic
oscillator in the cyclic representation with special non diagonal boundary conditions.
Moreover, the spectrum and Bethe states of the lattice versions of
the sine-Gordon and Liouville models with open diagonal boundaries
is examined. The role of the conserved quantities (boundary
non-local charges) in the derivation of the spectrum is also
discussed.

\end{abstract}

\vfill


\baselineskip=16pt
\end{titlepage}

\section{Introduction}

The ultimate goal when studying a physical system is the
derivation of the corresponding observables. In the framework of
integrable spin chains the first step towards such an aim is the
diagonalization of the corresponding transfer matrix by means of
the quantum inverse scattering method, introduced by the St.
Petersburg school \cite{FTS, FT, korepin}. The diagonalization
process relies primarily on the existence of certain exchange
relations emerging from the specific algebra that rules the model,
defined by either the Yang-Baxter \cite{korepin, baxter, tak} or
the reflection equation \cite{cherednik, sklyanin}. From the
physical viewpoint Yang--Baxter and reflection equations describe
the factorization of multiparticle scattering in the whole and
half line respectively, a unique feature displayed by 2D
integrable systems. Due to analyticity requirements imposed upon
the spectrum certain constraints, known as Bethe ansatz equations
\cite{FT, bethe} arise, whose exact form depends explicitly on the
choice of representation. The significance of Bethe ansatz
equations rests on the fact that their solutions yield all the
physically relevant quantities such as exact $S$ matrices
\cite{FT, rewi, done}, thermodynamic properties \cite{t1, jm, ftw}
and correlation functions \cite{korepin, maillet}. In addition it
was recently realized \cite{miza} that in a particular limit
string theories may be mapped to ${\cal N}=4$ super Yang-Mills
gauge theories, which in turn can be associated to special
examples of integrable spin chains. Thus one may implement the
powerful Bethe ansatz techniques \cite{FT, bethe}, in order to
derive non-perturbatively the physically relevant quantities. This
remarkable realization put integrable spin chains and Bethe ansatz
into the fore of recent advances in string theory.

The main aim of the present article is the investigation of the
spectrum of a generalized XXZ model, and associated models, with
special open boundary conditions that preserve integrability.
Historically the spin ${1\over 2}$ XXZ chain with diagonal
boundaries was investigated in \cite{sklyanin, gaudin, alcaraz}, whereas the
corresponding spin chain with non diagonal boundaries was just
recently solved for $q$ root of unity \cite{nepo}, and for
generic values of $q$ in \cite{chin}. The spectrum and Bethe
ansatz for the spin ${\mathrm s}$ XXZ model with non-diagonal
boundaries were also derived in \cite{doikouf, mart}. In
\cite{doma1} the spectral equivalence between the Hamiltonians of
the XXZ model with a non diagonal boundary and a novel open model (the asymmetric twin chain)
with an obvious reference state is shown via the representation
theory of boundary Temperley--Lieb algebras. Furthermore, the
spectral equivalence of XXZ type Hamiltonians with diagonal and
non-diagonal boundaries was investigated in \cite{nic}, again in
the context of boundary Temperley-Lieb algebras. Equivalence of
the spectra of spin chains with different boundaries was also
established in \cite{bajn} by introducing appropriate defects.
In the present study we shall be mainly interested in the case where the parameter $q$
is a root of unity, and we shall restrict our attention in finding
the spectrum of the general XXZ model and the $q$ harmonic
oscillator with non diagonal boundaries for the so called cyclic
representation \cite{roar}, which has not been treated so far. It
will be instructive to provide a special example of the
Hamiltonian of the spin ${1 \over 2}$ XXZ model with special
integrable conditions, i.e. \be {\cal H} = -{1\over 4}
\sum_{i=1}^{N-1}\Big (\sigma_{i}^{x}\sigma_{i+1}^{x}
+\sigma_{i}^{y}\sigma_{i+1}^{y}+ \cosh (i\mu)\
\sigma_{i}^{z}\sigma_{i+1}^{z}\Big )  -{N\over 4}\ \cosh (i\mu)
+{\cal M}_{N}  +c_{1}\ \sigma_{1}^{z}+c_{2}\ \sigma_{1}^{+}
+c_{3}\ \sigma_{1}^- \label{H0} \ee $\sigma^{x,y,z},\
(\sigma^{\pm})$ are the $2 \times 2$ Pauli matrices. The constants
$c_{i}$ are dictated by integrability, and ${\cal M}$ is a $2
\times 2$ matrix with entries being either scalars ($c$-number $K$
matrix applied at the right boundary) or elements of a particular
algebra, e.g. ${\cal U}_{q}(sl_{2})$, the $q$ harmonic oscillator
algebra etc., that is a dynamical boundary coupled to the right
end of the spin chain, that is a typical quantum impurity problem.
In general such type of Hamiltonians turn out to be of particular
significance in condensed matter physics, describing for instance
quantum impurities (see \cite{sklyanin, ftw, doikouf, bajn},
\cite{zab}--\cite{bako} and references therein) or being related
to the Azbel-Hofstadter Hamiltonian \cite{zab, wiza, faka, flni}.
Note that we shall also examine the
spectrum and Bethe ansatz equations of the open lattice sine
Gordon and Liouville models, but only with diagonal boundaries.

Let us outline the content of each section of the present article.
Sections 2 and 3 serve basically as a general review. More
precisely, in section 2 we introduce the generalized XXZ model,
and we also review how associated models such as the lattice
sine-Gordon, Liouville models and the $q$ harmonic oscillator
arise via certain limiting processes. The underlying algebraic
description is also provided and useful realizations are
presented. In fact, one of the the main objectives of this article
is to offer a unifying framework for the study of a whole class of open
integrable lattice models emerging from the generalized XXZ model. However
to identify the spectrum of the relevant models one has to fix a
representation and this is done in sections 6 and 7. In section 3
section well known $c$-number $K$ matrices, solutions of the
reflection equation are reviewed, and the algebraic transfer
matrix of an open spin chain is also recalled.

In section 4 a special example of a XXZ Hamiltonian (\ref{H0}),
with spin ${1\over 2}$ in the bulk coupled to some dynamical
system at the one end is considered.  More precisely, explicit
novel expressions of the dynamical boundary term ${\cal M}$ of the
Hamiltonian (\ref{H0}) are provided. Note that these expressions
are generic (algebraic) at this stage, i.e. they are independent
of the choice of representation for both the ${\cal U}_{q}(sl_{2})$
and the $q$ oscillator algebras, which are our primary interest in
the present study. In fact such a Hamiltonian may be regarded as a
special example of the more general case described in section 3.
To deal with the corresponding spectrum naturally one has to
choose a particular representation associated to the boundary.
This is done however in section 6, where we restrict our attention
to the cyclic representation for both XXZ and the $q$ harmonic
oscillator. In section 5 the boundary non local charges for the
generic XXZ chain are reviewed, and  novel expressions of the
non-local charges for the lattice sine-Gordon model and $q$
harmonic oscillator are derived. Their significance and their
relation to the spectrum becomes clear in section 6.

In section 6 the spectrum of the XXZ model and the $q$ harmonic
oscillator in the cyclic representation \cite{roar}, for special
non-diagonal boundaries is investigated. Generalizing the
formulation of \cite{chin, FT2} we are able to identify a
reference state, by means of suitable local gauge transformations
known also as Darboux matrices \cite{sklyanind}, upon which
eigenstates are built. The derivation of the reference states,
which is a new result, is achieved by solving sets of recursion
relations. Within the algebraic Bethe ansatz framework novel
expressions of the spectrum and Bethe ansatz equations of the
relevant models are found. In this context the Hamiltonian
presented in section 4, with the boundary terms associated to the
cyclic representation, (see also (\ref{H0})) is also treated as a
special case. Another intriguing new result is the connection
between the spectrum of the open transfer matrix and the spectrum
of the conserved boundary non local charges, which is presented in
section 6.3. Although this is a generic result, which hold for any
representation we provide particular examples for the XXZ model
and the $q$ harmonic oscillator in the cyclic representation,
given that these models are our primary concern. In Appendix C we present such a conserved quantity in the spin ${\mathrm s}$
representation (locally) as a tridiagonal (Jacobi) matrix
\cite{sklyanin2}, and in the cyclic representation of ${\cal
U}_{q}(sl_{2})$, and we attempt its diagonalization. In section 7
we give a flavor on the lattice sine-Gordon and Liouville theories
with diagonal boundary conditions. This is the first time to our
knowledge that these models with open boundary conditions are
considered. Finally, in the conclusion section we summarize the
results of the present article.

\section{The XXZ and associated models}

In this section the XXZ model is introduced, and various closely
related models, which arise from it  naturally are reviewed. The
XXZ spin ${1 \over 2}$ $R$ matrix, associated to the fundamental
representation of ${\cal U}_{q} (sl_{2})$, acting on $({\mathbb
C}^{2})^{\otimes 2}$. In general, the $R$ matrix is a quantity
proportional to the physical $S$ matrix, acts on ${\mathbb
V}^{\otimes 2}$, and is a solution of the Yang-Baxter equation
\cite{korepin, baxter} \be R_{12}(\lambda_{1}-\lambda_{2})\
R_{13}(\lambda_{1})\ R_{23}(\lambda_{2}) =R_{23}(\lambda_{2})\
R_{13}(\lambda_{1})\ R_{12}(\lambda_{1}-\lambda_{2}), \label{YBE}
\ee which acts on ${\mathbb V}^{\otimes 3}$, and as customary
$~R_{12} = R \otimes {\mathbb I}$, $~R_{23} ={\mathbb I} \otimes
R$. The spin ${1\over 2}$ XXZ $R$ matrix in particular is given by
\be R(\lambda) = \left(
\begin{array}{cc}
\sinh\mu(\lambda +{i\over 2} +{i\sigma^z \over 2})   &\sigma^{-} \sinh (i \mu)  \\
\sigma^{+}  \sinh (i \mu)   &\sinh\mu(\lambda +{i \over 2}
-{i\sigma^z \over 2})
\end{array} \right). \label{r} \ee The $R$ matrix is written in the so called
principle gradation, it can be however expressed in the
homogeneous gradation by means of a simple gauge transformation
\be R_{12}^{(h)}(\lambda) = {\cal V}_{1}(-\lambda)\
R_{12}^{(p)}(\lambda)\ {\cal V}_{1}(\lambda), ~~~~~{\cal
V}(\lambda) = diag(1,\ e^{\mu\lambda}). \label{gauge} \ee One may
now derive a more general object $L(\lambda) \in
\mbox{End}({\mathbb C}^2) \otimes {\cal U}_{q}(\widehat{sl_{2}})$,
where the quantum algebra ${\cal U}_{q}(\widehat{sl_{2}})$ is
defined by the fundamental algebraic relation (see also Appendix
A) \be R_{ab}(\lambda_{1} -\lambda_{2})\ L_{an}(\lambda_{1})\
L_{bn}(\lambda_{2})= L_{bn}(\lambda_{2})\ L_{an}(\lambda_{1})\
R_{ab}(\lambda_{1} -\lambda_{2}). \label{funda}\ee A simple
solution of the latter equation, which we shall use hereafter, is
(index free notation) \be L(\lambda) = \left(
\begin{array}{cc}
e^{\mu\lambda} {\mathrm A}-e^{-\mu \lambda} {\mathrm D}    &(q-q^{-1}){\mathrm B}\\
(q-q^{-1}){\mathrm C}    &   e^{\mu \lambda} {\mathrm D}-e^{-\mu \lambda} {\mathrm A} \\
\end{array} \right)  \label{l1} \ee
${\mathrm A}$, ${\mathrm B}$, ${\mathrm C}$, ${\mathrm D}$ satisfy
the defining relations of ${\cal U}_{q}(sl_{2})$ ($q=e^{i\mu}$),
namely \be && {\mathrm A}\ {\mathrm D} ={\mathrm D}\ {\mathrm A}
={\mathbb I}, ~~~~{\mathrm A}\ {\mathrm C} =q{\mathrm C}\ {\mathrm
A},  ~~~~{\mathrm A}\ {\mathrm B}= q^{-1} {\mathrm B}\ {\mathrm
A},~~~~\Big [{\mathrm C},\ {\mathrm B} \Big ] = {{\mathrm A}^{2}
-{\mathrm D}^{2} \over q -q^{-1}}. \label{defin}  \ee A well known
representation of ${\cal U}_{q}(sl_2)$ is the spin ${\mathrm s}$
representation, which is ${\mathrm n}=2{\mathrm s}+1$ dimensional, and may
be expressed in terms of ${\mathrm n}\times {\mathrm n}$ matrices
as \be A = \sum_{k=1}^{{\mathrm n}} q^{\alpha_{k}}\ e_{kk}, ~~~~~C
=\sum_{k=1}^{{\mathrm n}-1} \tilde C_{k}\ e_{k\ k+1},
~~~~~B=\sum_{k=1}^{{\mathrm n}-1} \tilde C_{k}\ e_{k+1\ k}
\label{spins} \ee where we define the matrix elements
$(e_{ij})_{kl} = \delta_{ik}\ \delta_{jl}$ and \be \alpha_{k} =
{{\mathrm n}+1 \over 2} - k, ~~~~~\tilde C_{k} = \sqrt{[k]_q
[{\mathrm n}-k]_q}, ~~~~~[k]_q = {q^k - q^{-k} \over q -q^{-1}}.
\ee The generators ${\mathrm A}$, ${\mathrm B}$, ${\mathrm C}$,
${\mathrm D}$ may be also expressed in terms of the
Heisenberg-Weyl group generators ${\mathbb X},\ {\mathbb Y}$ \be {\mathbb X}\ {\mathbb Y} =q\ {\mathbb Y}\ {\mathbb X}
\label{he}\ee as \be  && {\mathrm A}= {\mathrm D}^{-1} = {\mathbb X},
~~~{\mathrm B} ={1 \over q-q^{-1}}(q^{-s}{\mathbb X}^{-1} - q^{s}{\mathbb X}){\mathbb Y}^{-1},
~~~{\mathrm C} ={1 \over q-q^{-1}}(q^{-s}{\mathbb X} - q^{s}{\mathbb X}^{-1}){\mathbb Y}.
\label{terms} \ee Such a realization may be also thought of as a
$q$-deformation of the $sl_{2}$ algebra, when the corresponding
generators are expressed in  terms of differential operators, i.e.
\be S^{z}= y{d\over d y},~~~~S^{-}= y^{-1}(y{d\over d y}+{\mathrm
s}),~~~~S^{+}= y(y{ d\over d y}-{\mathrm s}).  \ee It is worth
pointing out that the latter expressions for ${\mathrm s}=0$,
correspond to a $sl(2,{\mathbb R})$ realization, which has been
employed in the context of high energy QCD. More precisely, it was
shown in \cite{lip} that the high energy asymptotic behavior of
the hadron-hadron scattering amplitude in QCD is described by the
non-compact XXX Heisenberg chain with ${\mathrm s}=0$ \cite{kor,
lip2}.

In the present article the XXZ model will be examined for the
special case where $q$ is root of unity, i.e. $q^{p}=1,\
q=e^{i\mu},\ \mu={2k\pi \over p}$, where $k,\ p$ integers. In this
case the algebra admits a $p$ dimensional representation, known as
the cyclic representation \cite{roar}. More specifically, in
addition to (\ref{he}) one more restriction is applied so one may
obtain a representation with no highest (lowest) weight \be
{\mathbb X}^p={\mathbb Y}^p=1 \nonumber \ee then the generators ${\mathbb X},\ {\mathbb Y}$ may be
expressed as $p$ dimensional matrices \be {\mathbb X} = \sum_{k=1}^p q^{-k}\
e_{kk},~~~~ {\mathbb Y}=\sum_{k=1}^{p-1}e_{k\ k+1} + e_{p1}.\label{cyclic}
\ee This realization is of particular interest and it has been
extensively used for instance in the problem of Bloch electrons in
a magnetic field, described by the Azbel-Hofstadter Hamiltonian
\cite{wiza, faka, flni}. In addition as argued in  \cite{hall} the
quantum Hall effect states may be also seen as cyclic
representation of ${\cal U}_{q}(sl_{2})$.

The XXZ model can be regarded as a universal model, since it gives
rise to a whole class of associated integrable models. In what
follows we shall briefly review how the lattice sine-Gordon and
Liouville models are obtained in a natural way from the XXZ $L$
matrix. Also the $q$ harmonic realization will be obtained again
from the generalized XXZ form. The generators ${\mathbb X}$ and ${\mathbb Y}$ may be
associated with an infinite dimensional representation in terms of
some lattice `fields'. Then one may identify the $L$ operator of
the lattice sine--Gordon model \cite{izko}. More specifically set
${\mathbb Z}_{n} ={\mathbb Y}_{n}\ {\mathbb X}_{n}$, it is then clear that ${\mathbb X}_{n}\ {\mathbb Z}_{n} = q
{\mathbb Z}_{n}\ {\mathbb X}_{n} $, where notice that for convenience we reintroduce
the index $n$.  Also set \be {\mathbb X}_{n} \to e^{-i \Phi_{n}},
~~~~~{\mathbb Z}_{n} \to e^{i\Pi_{n}},~~~~ \Big[\Phi_{n},\ \Pi_{m} \Big] =
i\mu\delta_{mn}  \label{xy} \ee $q^{s-{1\over 2}} = -i m$. The parameter $s$
of the representation is associated to the mass scale of the
system. Also by multiplying by $-im\ \sigma^x$ (we are allowed to
multiply with $\sigma^{x}$ because this leaves the XXZ $R$ matrix
invariant) one obtains the lattice sine-Gordon $L$ matrix
\cite{izko} \be L_{an}^{SG}(\lambda) = \left(
\begin{array}{cc}
h_{+}(\Phi_{n})e^{i\Pi_{n}}   &-i m\ 2 \sinh(\mu \lambda+i\Phi_{n})\\
-im\ 2 \sinh(\mu \lambda-i\Phi_{n})  &   h_{-}(\Phi)e^{-i\Pi_{n}}  \\
\end{array} \right) \label{sg}  \ee
\be h_{\pm}(\Phi_{n}) = 1 +m^{2}e^{\pm 2i\Phi_{n} +i\mu}.
\nonumber \ee As was argued in \cite{fati} by taking an
appropriate massless limit a lattice version of the Liouville
model is also recovered. Indeed by multiplying the sine--Gordon
$L$ operator with \be L_{an}(\lambda) = g_{a}\
L_{an}^{SG}(\lambda)\ g_{a}^{-1},~~~~~~ g=diag \Big (({m \over i
\alpha} )^{{1\over 2}},\ ({m\over i \alpha })^{-{1\over 2}} \Big )
\label{gauge2} \ee where $\alpha$ will be associated to the
spacing of the lattice and in the classical limit $\alpha \to 0$
(see below). In the quantum level is a finite number and for our
purposes here we set $\alpha = -i$ (see also section 7). Consider
also the following limiting process \cite{fati} \be  i\Phi_{n} \to
i\Phi_{n} +c, ~~~~ \mu \lambda \to \mu \lambda +c, ~~~~m \to 0,
~~~~e^{-c} \to \infty, ~~~~m^2 e^{-2c} \to \alpha^{2} \label{lim}
\ee one obtains the lattice Liouville $L$ matrix \be
L_{an}^{Lv}(\lambda) = \left(
\begin{array}{cc}
e^{i \Pi_{n}}   &\alpha e^{-\mu \lambda -i\Phi_{n}}\\
2 \alpha \sinh(\mu \lambda -i\Phi_{n})&h(\Phi_{n})e^{-i\Pi_{n}}\\
\end{array}\right )  \label{liou} \ee
\be h(\Phi_{n})= 1 +\alpha^2 e^{-2i\Phi_{n} +i\mu}. \nonumber\ee
The interesting observation is that the entailed $L$ operator
(\ref{liou}) has a non trivial spectral ($\lambda$) dependence a
fact that allows the application of Bethe ansatz techniques for
the derivation of the spectrum (see also \cite{fati}). The
classical limit of the aforementioned $L$ matrices (\ref{sg}),
(\ref{liou}) provides the corresponding classical Lax operators
satisfying the zero curvature condition, and giving rise to the
classical equations of motion of the relevant models (for more
details see also \cite{izko, fati}). Let us briefly review the
connection between the quantum (lattice) versions and the
classical sine-Gordon and Liouville models. Consider the following
classical limit \cite{izko, fati}, the spacing $\alpha \to 0$, set
$\mu \to h \mu$ such that ${1 \over h}[ , ] \to \{ , \}$, and \be
\Phi_n \to {\beta \over 2} \phi(x) -{\pi \over 2}, ~~~~~\Pi_n \to
\alpha {\beta \over 4} \pi(x), ~~~~ -4im \to \alpha  \tilde m,
~~~\mu\lambda \to u +{i\pi \over 2}  \label{clas} \ee $\tilde m$
is the continuum mass and $\beta$ corresponds to the coupling
constant of the sine Gordon model, and for the Liouville model we
set ${\beta \over 2} =1$, following the normalization of
\cite{fati}. Bearing in mind the expressions above (\ref{clas}) we
obtain as $\alpha\to 0$ \be L(u) =1 - \alpha {\mathrm U}(u) +{\cal O}(\alpha^2) \ee then the
quantities ${\mathrm U}(u)$ and ${\mathrm V}(u)$, written below provide a Lax pair for
the classical counterparts of the lattice sine-Gordon and
Liouville models. More precisely for the sine Gordon model \be &&
{\mathrm U}(u) = {1\over 2} \left(
\begin{array}{cc}
-i {\beta \over 2}\pi(x)   &-\tilde m  \sinh(u+i{\beta \over 2}\phi(x))\\
\tilde m \sinh(u-i{\beta \over 2}\phi(x))  &   i{\beta \over 2}\pi(x)  \\
\end{array} \right),  \non\\ && {\mathrm V}(u) = {1\over 2} \left(
\begin{array}{cc}
-i {\beta \over 2} \phi'(x)   &-\tilde m  \cosh(u+i{\beta \over 2}\phi(x))\\
\tilde m \cosh(u-i{\beta \over 2} \phi(x))  &   i{\beta \over 2} \phi'(x)  \\
\end{array} \right) \label{sgc} \ee whereas for the Liouville
model the Lax pair reads \be  {\mathrm U}(u)= {1\over 2} \left(
\begin{array}{cc}
-i\pi(x)   & -2 e^{-u -i\phi(x)} \\
4 \sinh(u -i\phi(x)) &i\pi(x)\\
\end{array}\right ), ~~~~{\mathrm V}(u)= {1\over 2} \left(
\begin{array}{cc}
-i\phi'(x)   & 2 e^{-u -i\phi(x)} \\
4\cosh(u -i\phi(x)) &i\phi'(x)\\
\end{array}\right )  \label{liouc} \ee
The Lax pair satisfies the zero curvature condition \be \dot{{\mathrm U}} -
{\mathrm V}' + \Big [{\mathrm U},\ {\mathrm V} \Big ]=0  \ee which leads to the corresponding
equations of motions i.e. \be && \mbox{sine-Gordon model:}
~~~~~\ddot{\phi}(x) - \phi''(x) +{\tilde m^{2} \over \beta}
\sin(\beta \phi(x))=0 \non\\  && \mbox{Liouville model:}
~~~~~\ddot{\phi}(x) - \phi''(x) -4 i e^{-2i\phi(x)}=0. \ee

Similar limiting process to (\ref{lim}) leads to the $q$-harmonic
oscillator $L$ matrix starting from (\ref{l1}), (\ref{terms}). In
fact, by simply multiplying the Liouville $L$ matrix with an
anti-diagonal matrix and bearing also in mind (\ref{xy}) we obtain
the following \be L_{an}(\lambda) =\left(\begin{array}{cc}
e^{\mu \lambda}V_{n} - e^{-\mu \lambda}V_{n}^{-1} &a_{n}^{+} \\
       a_{n}                      &-e^{-\mu \lambda}V_{n} \\ \end{array} \right ) \label{l2} \ee where the operators
$a_{n},\ a_{n}^{+},\ V_{n}$ are expressed in terms of ${\mathbb X}_{n},\
{\mathbb Y}_{n}$ as \be V_{n}= {\mathbb X}_{n}, ~~~~a_n^+ = ({\mathbb X}_{n}^{-1}
-q{\mathbb X}_{n}){\mathbb Y}_{n}^{-1}, ~~~a_{n}= {\mathbb Y}_{n}\ {\mathbb X}_{n} \ee and they satisfy
the $q$ harmonic oscillator algebra i.e. \be a_{n}^{+}\ a_{n} =
1-q V_{n}^{2}, ~~~~a_{n}\ a_{n}^{+} = 1-q^{-1}V_{n}^2.
\label{qalg} \ee The latter may be also seen as $q$ deformation of
the Discrete-Self-Trapping (DST) model \cite{pronko}. The DST
equation introduced in \cite{dst} to describe the non-linear
dynamical behaviour of small molecules (ammonia, acetylene), and
big molecules as well (acetanilide). Also the integrability
properties of the DST model with two or more degrees of freedom
were studied in a series of articles \cite{sklyanind, dst2}.

Finally, the simplest $q$ deformed $L$ matrix is obtained from
(\ref{sg}) ---written in terms on the algebraic objects $X_{n},\
Z_{n}$ (see (\ref{xy}))--- via a simple limit \be \mu \lambda \to
\mu \lambda + b, ~~~~~e^{b} \to \infty, ~~~~~-i m e^{b} \to 1\ee
then the corresponding $L$ matrix becomes (see also \cite{faka})
\be L_{an}(\lambda) =\left(\begin{array}{cc}
{\mathbb Z}_{n}  & e^{\mu \lambda} {\mathbb X}_{n}^{-1} \\
e^{\mu \lambda} {\mathbb X}_{n}  &{\mathbb Z}_{n}^{-1}\\ \end{array} \right ) \ee and
this $L$ matrix was used for the construction of lattice versions of
the KdV system (for more details on the lattice KdV see also \cite{volkov, kudu, fioro}).

\section{The open transfer matrix}

As is well known
to construct the transfer matrix associated to an open spin chain one needs in addition
to the $R$ ($L$) matrix one more fundamental object, namely the
$K$ matrix, which is proportional to the physical boundary $S$
matrix, acting on ${\mathbb V}$, and is a solution of the
reflection equation \cite{cherednik}, \begin{equation}
R_{12}(\lambda_{1}-\lambda_{2})\ K_{1}(\lambda_{1})\
R_{21}(\lambda_{1}+\lambda_{2})\ K_{2}(\lambda_{2})=
K_{2}(\lambda_{2})\ R_{12}(\lambda_{1}+\lambda_{2})\
K_{1}(\lambda_{1})\ R_{21}(\lambda_{1}-\lambda_{2}). \label{re}
\end{equation} As proposed in \cite{doma, male} an effective way
of finding solutions of the reflection equations (\ref{re}) is to
exploit certain algebraic structures such as (boundary)
Temperley--Lieb algebras \cite{male}--\cite{mawo}.

Let us recall how one may express the XXZ $R$ and $K$ matrices,
solutions of the Yang-Baxter and reflection equations
respectively, in terms of the generators of the boundary
Temperley-Lieb (blob) algebra. First define the blob algebra
$b_N(q,Q)$, which is a quotient of the affine Hecke algebra
\cite{cher}, with generators ${\mathbb U}_{1},{\mathbb
U}_{2},...,{\mathbb U}_{N-1}$ and ${\mathbb U}_{0}$, and
relations:
\begin{eqnarray} {\mathbb U}_{l}\ {\mathbb U}_{l} &=& \delta\ {\mathbb U}_{l},~~~~{\mathbb U}_{0}\
{\mathbb U}_{0} = \delta_{0}\ {\mathbb U}_{0} \non\\
{\mathbb U}_{l\pm 1}\ {\mathbb U}_{l}\ {\mathbb U}_{l\pm 1} &=&
{\mathbb U}_{l\pm 1}, ~~~~{\mathbb
U}_{1}\ {\mathbb U}_{0}\ {\mathbb U}_{1} = \gamma\ {\mathbb U}_{1} \non\\
\Big [ {\mathbb U}_{l},\ {\mathbb U}_{k} \Big ] &=& 0, ~~~~|l-k|
\neq 1 \label{TL}
\end{eqnarray} $\delta=-(q+q^{-1})$, $q=e^{i\mu}$, and $\delta_{0}$, $\gamma$ are constants depending on $q$ and
$Q=e^{i{\mathrm m} \mu}$. Physically the constants $q$ and $Q$
play the role of the bulk and boundary parameter respectively of
the open spin chain, which will be constructed in the following.

The generators ${\mathbb U}_{l},\ l\in \{1,..,N-1 \}$, ${\mathbb
U}_{0}$ of the blob algebra are given in the spin ${1\over 2}$ XXZ
representation, i.e. let the tensor representation $h: b_{N}(q,Q)
\to \mbox{End}((\mathbb C^{2})^{\otimes N})$ such that (see also
\cite{masa}) \be h({\mathbb U}_{l}) &=& 1 \otimes \ldots \otimes
\left(
\begin{array}{cccc}
    0    &0        &0       &0   \\
    0    &-q       & 1      &0   \\
    0    &1        &-q^{-1} &0   \\
    0    &0        &0       &0
\end{array} \right) \otimes \ldots \otimes 1,\non\\ h({\mathbb U}_{0}) &=&\left( \begin{array}{cc}
      -Q      & q^{ \theta}       \\
       q^{-\theta}      &-Q^{-1}    \\
\end{array} \right)
\otimes \ldots \otimes 1 \label{tlg} \ee with $h({\mathbb U}_{l})$
acting non-trivially on ${\mathbb V}_{l} \otimes {\mathbb
V}_{l+1}$, (where ${\mathbb V} =\mathbb C^{2}$) and $h({\mathbb
U}_{0})$ acting on ${\mathbb V}_{1}$. Note that $h({\mathbb
U}_{l})$ $l\in \{1, \ldots, N-1 \}$ are actually the XXZ
representations of the Temperley-Lieb algebra generators \cite{tl,
mawo}. The notation here is slightly modified compared to
\cite{doma}. Also, for this representation we find: $\delta_{0} = -(Q+Q^{-1})$ and $\gamma = q Q +q^{-1}Q^{-1}$.

As argued in \cite{doma} tensor representations of the blob
algebra provide solutions of the reflection equation. Hence a
solution of the reflection equation (\ref{re}) may be written in
terms of the blob algebra generator $h({\mathbb U}_{0})$
(\ref{tlg}) as \be K_1^{(b)}(\lambda) = 2\sinh\mu (\lambda-{i
{\mathrm m} \over 2}-i\zeta)\ \cosh\mu (\lambda-{i {\mathrm m}
\over 2}+i\zeta)\ {\mathbb I} + \sinh (2\mu\lambda)\ h({\mathbb
U}_{0}), \label{ansatz1} \ee $\zeta$ is an arbitrary constant. The
XXZ spin ${1 \over 2}$ $R$ matrix (\ref{r}) may be also written in
terms of the representation $h({\mathbb U}_{l})$, \cite{jimbo},
i.e. \be R_{12}(\lambda) = {\cal P}_{12}\Big (\sinh \mu(\lambda
+i)\ {\mathbb I} +\sinh(\mu \lambda)\ h({\mathbb U}_{1}) \Big )
\label{ansatz2} \ee where ${\cal P}$ is the permutation operator,
such that ${\cal P}(a\otimes b)= b \otimes a$.

The matrix obtained from the blob algebra (\ref{ansatz1})
coincides with the $K$ matrix found in \cite{DVGR, GZ} (written in
the homogeneous gradation) i.e. \be  K(\lambda) =  \left(
\begin{array}{cc}
\sinh \mu(-\lambda +i\xi)\ e^{\mu \lambda}  &q^{\theta} \kappa \sinh (2\mu\lambda)         \\
q^{-\theta} \kappa \sinh(2\mu \lambda)      &\sinh \mu(\lambda
+i\xi)\ e^{-\mu \lambda}
\end{array} \right) \label{kc} \ee
subject to the following identifications \be {e^{-i\mu \xi} \over
2\kappa} = \sinh (i {\mathrm m} \mu), ~~~{e^{i\mu \xi} \over
2\kappa}= -\sinh (2i \mu \zeta). \label{ide} \ee The $K$
matrix (\ref{ansatz1}), (\ref{kc}) is written in the homogeneous
gradation. To obtain the matrix in the principal gradation it is
necessary to perform a simple gauge transformation, i.e. \be
K^{(p)}(\lambda)= {\cal V}(\lambda)\ K^{(h)}(\lambda)\ {\cal
V}(\lambda) \ee where ${\cal V}$ is given by (\ref{gauge}).

Another representation that leads eventually to upper (lower)
triangular $K$ matrices follows essentially if one considers the
limit $q^{\theta},\ Q \to \infty$, then the representation of the
boundary element (\ref{tlg}) reduces to: \be h({\mathbb U}_{0})
&=&\left( \begin{array}{cc}
      -Q      & q^{ \theta}       \\
       0      &0\\
\end{array} \right)
\otimes \ldots \otimes 1. \label{tlg2} \ee Indeed the matrices
$h({\mathbb U}_{l})$ (\ref{tlg}) and $h({\mathbb U}_{0})$
(\ref{tlg2}) form a representation of the blob algebra with
$\delta_{0}=-Q$ and $\gamma  =q\ Q$. The corresponding $K$ matrix
then is \be K(\lambda) =  \left(
\begin{array}{cc}
\sinh \mu(-\lambda +i\xi)\ e^{\mu \lambda}  &q^{\theta} \kappa \sinh (2\mu\lambda) \\
0      &\sinh \mu(\lambda +i\xi)\ e^{-\mu \lambda}
\end{array} \right) \label{kk} \ee where in this case the relations among the boundary parameters become
\be {e^{-i\mu \xi} \over \kappa} = e^{i {\mathrm m} \mu},
~~~{e^{i\mu \xi} \over 2\kappa}= -\sinh (2i \mu \zeta).
\label{ide2} \ee It is clear that by considering $q^{-\theta} \to
\infty$ one recovers a lower triangular matrix. In fact, such
upper (lower) triangular $K$ matrices may be thought of as the
result of a very particular gauge transformation of the type
(\ref{gauge2}), where the mass parameter $m\propto q^{\theta}$
becomes very big (or very small).

Finally, to construct the generating function of the conserved
commuting quantities associated to an open spin chain, one needs two
solutions $K^\pm(\lambda)$ of the reflection equation (\ref{re}).
Indeed, the transfer matrix $t(\lambda)$ associated to an open
spin chain of $N$ sites \cite{sklyanin} reads
\be t(\lambda) = \tr_{0}  \Big \{ K^{+}_{0}(\lambda)\  T_{0}(\lambda)\
K^{-}_{0}(\lambda)\ \hat T_{0}(\lambda) \Big \}\ = \tr_{0} \Big \{
K_{0}^{+}(\lambda)\ {\cal T}_{0}(\lambda) \Big \} \label{transfer2} \ee
where ${\cal T}$ is a solution of the reflection equation
(\ref{re}) and
\be  T_{0}(\lambda) = L_{01}(\lambda-i\Theta)
L_{02}(\lambda-i\Theta) \cdots L_{0N}(\lambda-i\Theta)\, ~~~\hat
T_{0}(\lambda) = \hat L_{0N}(\lambda +i\Theta) \cdots \hat
L_{02}(\lambda+i\Theta) \hat L_{01}(\lambda+i\Theta)
\label{hatmonodromy}, \ee $L$ is one of the matrices introduced in section 2 and $\hat L(\lambda)
=L^{-1}(-\lambda)$.  As usually we keep only the index $0$,
associated to the `auxiliary' space, whereas the indices $1,
\ldots, N$, corresponding to the `quantum' spaces are suppressed.
The parameter $\Theta$ is called inhomogeneity, and in principle
one could attach a different one at each site of the spin chain.
Also $K^{+}(\lambda) =M\ K(-\lambda -i)$, $K$ is
any solution of the reflection equation, and $M$ is defined as:
\be && M= {\mathbb I}, ~~~~~~~~~~~~~~~~~\mbox{Principal gradation}
\non\\ && M =diag(q,\ q^{-1}), ~~\mbox{Homogeneous gradation}. \ee
$K^-$ is a $c$-number solution (\ref{kc}) of the reflection
equation as well. The transfer matrix (\ref{transfer2}) provides a family
of commuting operators \cite{sklyanin}, i.e. \be \Big
[t(\lambda),\ t(\lambda')\Big ]=0. \label{comt} \ee At this stage
the quantity (\ref{transfer2}), (\ref{hatmonodromy}) is a purely
algebraic construction. It only acquires a physical meaning
describing a spin chain system once particular representations are
chosen to act on the quantum spaces.

\section{Hamiltonian of a spin chain coupled to a dynamical boundary}

Already in the introduction the general form of a Hamiltonian with
dynamical boundary term was presented (\ref{H0}). We focus here in
a special case of the algebraic transfer matrix (\ref{transfer2}),
(\ref{hatmonodromy}) described in the previous section. More
precisely, we consider the situation where the spin ${1\over 2}$
representation is assigned at each quantum space ($L \mapsto R$, set
also $\Theta =0$), and the right end of the chain is coupled to
dynamical degrees of freedom, associated to the ${\cal
U}_{q}(sl_{2})$, and the $q$ harmonic oscillator algebras i.e. \be
t(\lambda) = tr_{0} \Big \{ K_{0}^{+}(\lambda)\ R_{01}(\lambda)
\ldots R_{0N}(\lambda )\ K_0^-(\lambda)\ \hat R_{0N}(\lambda) \ldots
\hat R_{01}(\lambda ) \Big \} \ee where $R$ is the XXZ matrix (\ref{r}), and now the $K^-$ matrix is
expressed in a factorized form \cite{sklyanin, zab, lf, doikouf,
doma, bako, bajn} \be K_{0}^{-}(\lambda) = L_{0{\mathrm
d}}(\lambda -i\Theta_{0})\ K_{0}(\lambda)\ \hat L_{0{\mathrm
d}}(\lambda+i\Theta_{0}) \label{kd} \ee the index ${\mathrm d}$
characterizes the extra `dynamical space' attached to the right
boundary of the chain, $K$ is a $c$-number solution of the
reflection equation (\ref{kc}) with parameters $\xi^-,\ \kappa^-,\
\theta^-$, $L$ being the matrix given by (\ref{l1}) or (\ref{l2}),
and $K^+ =M K'(-\lambda -i)$ where $K'$ is again a $c$-number
solution of the reflection equation of the general type
(\ref{kc}), but with parameters $\xi^+,\ \kappa^+,\ \theta^+$.
Note that the dynamical space may be thought of as an extra site
of the spin chain, but here it is preferable to treat it as a
dynamical system attached to the boundary. It will be instructive
for future reference to provide at this point explicit expressions
for the dynamical matrix ${\cal M}$ appearing in the Hamiltonian
(\ref{H0}), and due to the dynamical nature of the boundary the
entries of ${\cal M}$ in (\ref{H0}) will be written in terms
of the generators of the ${\cal U}_{q}(sl_{2})$ and  $q$ harmonic
oscillator algebras as we will see below.  In general the
Hamiltonian is defined as proportional to ${d \over d \lambda}
t(\lambda)\vert_{\lambda =0}$, in particular we choose to
normalize as (we also set $\Theta =0$): \be {\cal H} = -{(\sinh
(i\mu))^{-2N+1}{\cal F}^{-1} \over {4 \mu \sinh(i\mu \xi^-)}} \Big( tr_{0}
K_{0}^{+}(\lambda) \Big )^{-1}\ \ {d \over d \lambda}
t(\lambda)\vert_{\lambda =0} \label{H01}, \ee where ${\cal F}= L_{0d}(-i\Theta_0 )\ \hat L_{0d}(i \Theta_0)$, and for:
\\
\\
{\bf (I)} the ${\cal U}_{q}(sl_{2})$ case, ${\cal F} = w-\cosh
i\mu(1+2\Theta_{0})$ with $w$ being the Casimir of ${\cal
U}_q(sl_2)$ \be w = q\ A^2 +q^{-1} D^2+(q-q^{-1})^2B\ C = q^{-1}\
A^2 +qD^2 + (q-q^{-1})^2 C\ B. \ee {\bf (II)} the $q$-harmonic
oscillator, ${\cal F} = 1 -q^{1+2\Theta_0}$.

Then having in mind
that  $~R(0)= \hat R(0)=\sinh i \mu \ {\cal P}, ~~~K(0) = \sinh
(i\mu \xi^-)\ {\mathbb I}~$ and also define \be H_{kl}= -{1\over
2 \mu}{d\over d\lambda}({\cal P}_{kl}\ R_{kl}(\lambda)) \ee we may
rewrite the Hamiltonian as \be {\cal H} = \sum_{l=1}^{N-1} H_{l\
l+1} - {\sinh (i\mu)\ {\cal F}^{-1} \over 4 \mu \sinh(i\mu \xi^-)}  \left( {d
\over d \lambda} K^-_{N}(\lambda) \right) \Big \vert_{\lambda =0}
+ {tr_{0}\ K_{0}^{+}(0)\ H_{1 0} \over tr_{ 0}\ K_{0}^{+}(0)}.
\label{H} \ee The first term in (\ref{H}) gives rise to the bulk
spin-spin interaction between first neighbours appearing in
(\ref{H0}), whereas the second term gives rise to ${\cal M}$, and
the third one to the last three terms of (\ref{H0}). We shall
provide at this point the explicit values of the constants $c_{i}$
of (\ref{H0}). It is straightforward to compute from (\ref{H})
that: \be c_{1}= - {\sinh (i\mu) \cosh (i\mu \xi^{+}) \over 4
\sinh (i\mu \xi^{+})}, ~~~~~c_{2}=- {\sinh (i\mu)
\kappa^{+}q^{-\theta^+} \over 4 \sinh (i\mu \xi^{+})}, ~~~~~c_{3}
=- {\sinh (i\mu) \kappa^{+}q^{\theta^+} \over 4 \sinh (i\mu
\xi^{+})}, \ee where the parameters $\xi^+,\ \kappa^+,\ \theta^+$
are apparently associated to the left boundary $K^+$. To summarize
the Hamiltonian is finally written exactly as in (\ref{H0}) \be
{\cal H} &=& -{1\over 4} \sum_{i=1}^{N-1}\Big
(\sigma_{i}^{x}\sigma_{i+1}^{x} +\sigma_{i}^{y}\sigma_{i+1}^{y}+
\cosh (i\mu)\ \sigma_{i}^{z}\sigma_{i+1}^{z}\Big )  -{N \over 4}\
\cosh (i\mu) +{\cal M}_{N}  \non\\ &-& {\sinh (i\mu) \cosh (i\mu
\xi^{+}) \over 4 \sinh (i\mu \xi^{+})} \sigma_{1}^{z} - {\sinh
(i\mu) \kappa^{+}q^{-\theta^+} \over 4 \sinh (i\mu \xi^{+})}
\sigma_{1}^{+} - {\sinh (i\mu) \kappa^{+}q^{\theta^+} \over 4
\sinh (i\mu \xi^{+})} \sigma_{1}^- \label{H1} \ee Our objective in
what follows is to give explicit expressions for the matrix ${\cal
M}$ occurring as a boundary contribution in the Hamiltonian above
due to the second term in (\ref{H}).
\\
\\
{\bf (I)} {\it The ${\cal U}_{q}(sl_{2})$ algebra}: The entries of
${\cal M}$ `act' to the `dynamical space' which in this case is
associated to a copy of ${\cal U}_{q}(sl_{2})$, hence they are expressed in terms of the ${\cal
U}_{q}(sl_{2})$ generators as \be {\cal M}_{11} &=& -{\sinh (i\mu
)\ {\cal F}^{-1} \over 4\sinh (i\mu \xi^-) }\Big[- \cosh (i\mu \xi^-)\Big ( 2q\
A^2 + 2q^{-1}D^2 -w -q^{1+2\Theta_0}-q^{-1-2\Theta_0}\Big ) \non\\
&+& 2 \kappa^- (q-q^{-1}) \Big (q^{-\theta^-}B(q^{1+\Theta_{0}} A-
q^{-1-\Theta_{0}}
D)+q^{\theta^-}(q^{-\Theta_{0}}A-q^{\Theta_{0}}D)C \Big ) + 2
\sinh (i\mu \xi^-) \Big (qA^2 -q^{-1} D^2 \Big )\Big ] \non\\
{\cal M}_{22}&=& -{\sinh (i\mu )\ {\cal F}^{-1} \over 4\sinh (i\mu \xi^-) }\Big [
\cosh (i\mu \xi^-)\Big (2 q^{-1}\ A^2 + 2 qD^2 -w
-q^{1+2\Theta_{0}}-q^{-1-2\Theta_{0}}\Big )  \non\\ &+&2 \kappa^-
(q-q^{-1}) \Big (q^{\theta^-}C(q^{1+\Theta_{0}} D-
q^{-1-\Theta_{0}}
A)+q^{-\theta^-}(q^{-\Theta_{0}}D-q^{\Theta_{0}}A)B \Big )+ 2 \sinh (i\mu
\xi^-) \Big (qD^2 -q^{-1} A^2 \Big ) \Big ]
\non\\ {\cal M}_{12} &=& -{\sinh (i\mu )\ {\cal F}^{-1} \over 4\sinh (i\mu \xi^-)
}\Big[ 2\cosh (i\mu
\xi^-)(q-q^{-1})(q^{\Theta_{0}}D-q^{-\Theta_{0}}A)B
+2\sinh(i\mu \xi^-) (q-q^{-1})(q^{-\Theta_0}A+q^{\Theta_0}D)B \non\\
&+& 2\kappa^- \Big (q^{-\theta^-}(q-q^{-1})^2 B^2
+q^{\theta^-}(q+q^{-1}-q^{1+2\Theta_0}D^2-q^{-1-2\Theta_0}A^2)
\Big) \Big ]\non\\ {\cal M}_{21} &=& -{\sinh (i\mu )\ {\cal F}^{-1} \over 4\sinh
(i\mu \xi^-) }\Big[ 2\cosh (i\mu
\xi^-)(q-q^{-1})(q^{-\Theta_0}D-q^{\Theta_0}A)C  +2\sinh(i\mu
\xi^-) (q-q^{-1})(q^{\Theta_0}A+q^{-\Theta_0}D)C \non\\ &+&
2\kappa^- \Big (q^{\theta^-}(q-q^{-1})^2 C^2
+q^{-\theta^-}(q+q^{-1}-q^{-1-2\Theta_0} D^2-q^{1+2\Theta_0}A^2)
\Big) \Big ].  \label{mm1}\ee
\\
{\bf (II)} {\it The $q$ harmonic oscillator algebra}: In this case
the entries of ${\cal M}$ belong to the $q$ harmonic oscillator
algebra (\ref{qalg}) generated by $V,\ a,\ a^+$ \be {\cal M}_{11}
&=& -{\sinh (i\mu )\ {\cal F}^{-1} \over 4\sinh (i\mu \xi^-) }\Big[ \cosh(i\mu
\xi^- )(1+q^{1 + 2 \Theta_0}-2qV^2) +2\sinh (i\mu \xi^-) qV^2
\non\\ && +2\kappa^- \Big (q^{-\theta^-+1}a^+V+q^{\theta^-}
(q^{-\Theta_0}V-q^{\Theta_0}V^{-1})a \Big )\Big ] \non\\ {\cal
M}_{22} &=&  -{\sinh (i\mu )\ {\cal F}^{-1} \over 4\sinh (i\mu \xi^-) }\Big[
-\cosh(i\mu \xi^- )(1+q^{1+2\Theta_0} -2q^{-1}V^2) -2 \sinh (i\mu
\xi^-)q^{-1} V^2 \non\\ && +2\kappa^- \Big
(-q^{-\theta^-}q^{\Theta_0}Va^+ - q^{\theta^-}
a(q^{-1-2\Theta_0}V-q^{1+2 \Theta_0}V^{-1}) \Big  ) \Big ] \non\\
{\cal M}_{12} &=&  -{\sinh (i\mu )\ {\cal F}^{-1}\over 4\sinh (i\mu \xi^-)
}\Big[ -2\cosh(i\mu \xi^-) (q^{-\Theta_0}V-q^{\Theta_0}V^{-1})a^+
+2\sinh (i\mu \xi^-)(q^{-\Theta_0}V+q^{\Theta_0}V^{-1})a^+ \non\\
&& +2\kappa^- \Big  (q^{\theta^-} (q+q^{-1}- q^{1+2\Theta_0}V^{-2}
-q^{-1-2\Theta_0}V^2) +q^{-\theta^-} (a^+)^2 \Big )\Big ] \non\\
{\cal M}_{21} &=&  -{\sinh (i\mu )\ {\cal F}^{-1} \over 4 \sinh (i\mu \xi^-)
}\Big[ -2\cosh(i\mu \xi^-) q^{\Theta_0}Va +2\sinh (i\mu \xi^-)
q^{\Theta_0}Va +2 \kappa^- \Big  (q^{\theta^-} a^2
-q^{-\theta^-}q^{-1+2\Theta_0}V^2)\Big ) \Big ]. \non\\
\label{mm2} \ee As expected the entries of the matrix ${\cal M}$,
due to right dynamical boundary, are expressed in
terms of the generators of the two main algebras i.e. ${\cal
U}_{q}(sl_{2})$ and the $q$ harmonic oscillator. These generators
occur naturally in the expressions above due to the presence of
$L_{0d}, \ \hat L_{0d}$ in the dynamical boundary we are
considering (\ref{kd}). To find the spectrum and the eigenstates
of such a Hamiltonian one has to fix the representation at the
boundary, and this will be done in section 6 where the
cyclic representation (\ref{cyclic}) is considered for both ${\cal
U}_{q}(sl_{2})$ and the $q$ harmonic oscillator.

\section{The boundary non-local charges}

The so called boundary non-local charges \cite{mene2}--\cite{base}
will be now introduced. Such charges are going to play an
essential role in the derivation of the spectrum as we shall see
in a subsequent section. They are obtained via the standard
procedure, by considering the asymptotics of the operator ${\cal
T}$ as $\mu \lambda \to \infty$ (see also \cite{dema, doikou1}).
Note that in what follows the $L,\ \hat L$ and $K^{\pm}$ matrices
are written in the principal gradation.
\\
\\
{\bf (I)}{\it The XXZ and lattice sine Gordon models}: The
asymptotics of $L(\lambda -\Theta),\ \hat L(\lambda +\Theta)$ as
$\mu \lambda \to \infty$, bearing in mind the evaluation
homomorphism (\ref{eval1}), is given as ($\mu \Theta$ is
considered finite) \be && L_{0n}(\mu \lambda \to \infty) \propto
\left(
\begin{array}{cc}
k_{1,n}  &    \\
&  k_{2,n}        \\
\end{array} \right) \ +  e^{-\mu \lambda} w\left(
\begin{array}{cc}
&  f_{1,n}  \\
 f_{2,n}  &         \\
\end{array} \right) + \ldots , \non\\ &&\hat L_{0n}(\mu \lambda \to \infty) \propto \left(
\begin{array}{cc}
k_{1,n}  &    \\
&  k_{2,n}        \\
\end{array} \right) +  e^{-\mu \lambda} \hat w  \left(
\begin{array}{cc}
&  e_{2,n}  \\
 e_{1,n}  &         \\
\end{array} \right) + \ldots \  \label{d0} \ee with $k_{2}=
k_{1}^{-1}$. The subscript
$n$ simply denotes that the corresponding objects are associated
to the $n^{th}$ site (`quantum space') of the chain,  \be w = 2\ \sinh (i\mu)\ q^{-{1\over 2}+\Theta'},\
~~~\hat w =  2\ \sinh (i\mu)\ q^{-{1\over 2}- \Theta'},\
~~~\Theta' =\Theta +{1 \over 2}. \ee By also considering $K^-$,
being of the form (\ref{kc}) with boundary parameter $\xi^-,\
\kappa^-, \theta^-$, as $\mu \lambda \to \infty$ (\ref{kc}) (the
constants $\xi^-,\ \kappa^-, \theta^-$ are considered to be
finite) \be K^-(\mu \lambda \to \infty)  \propto \left(
\begin{array}{cc}
       &  q^{\theta^-}\kappa^-   \\
       q^{-\theta^-}\kappa^- &    \\
\end{array} \right) +e^{-\mu \lambda} \left(
\begin{array}{cc}
-e^{-i\mu \xi^-}   &   \\
               &e^{i\mu \xi^-}     \\
\end{array} \right) + \ldots \, \label{f0} \ee
one may formulate the corresponding behavior of ${\cal T}$ (\ref{transfer2}), namely
\be {\cal T}(\mu \lambda \to \infty) \propto
\left(
\begin{array}{cc}
       & q^{\theta^-} \\
    q^{-\theta^-} &     \\
\end{array} \right) +2 \sinh (i\mu)\ e^{-\mu \lambda}\ \left(
\begin{array}{cc}
Q_{1}^{(N)} &  \\
                     &Q_{2}^{(N)}    \\
\end{array} \right) + \ldots \label{asym} \ee where the  non-local charges $Q_{i}^{(N)}$, $i\in\{1,\ 2 \}$ are given by the
following expressions \cite{mene2}--\cite{base} \be Q_{i}^{(N)} =
q^{-{1\over 2} -\Theta_{i}}K_{i}^{(N)}E_{i}^{(N)}+q^{{1\over
2}+\Theta_{i}} K_{i}^{(N)}F_{i}^{(N)}+x_{i} (K_{i}^{(N)})^{2},
~~~i\in \{ 1,\ 2\}. \label{Q} \ee $K_{i}^{(N)}$, $E_{i}^{(N)}$ and
$F_{i}^{(N)}$, provide $N$ coproducts of the quantum Kac--Moody
algebra ${\cal U}_{q}(\widehat{sl_{2}})$ (see also Appendix) \be
K_{i}^{(N)} = \Delta^{(N)}(k_{i}), ~~~~E_{i}^{(N)}
=\Delta^{(N)}(e_{i}),~~~~ F_{i}^{(N)} =\Delta^{(N)}(f_{i}),~~~~i
\in \{1,\ 2 \} \label{copp} \ee and the constants appearing in
(\ref{Q}) are: \be x_{1} = -{e^{-i \mu \xi^-} \over 2\kappa^-
\sinh (i \mu)},\ ~~~~~~x_{2} = {e^{ i \mu \xi^-} \over 2 \kappa^-
\sinh (i \mu)} , ~~~~\Theta_{1} = \Theta'
-\theta^-,~~~~\Theta_{2}=\Theta'+\theta^-. \ee It is clear that in
the case of an upper triangular $K^-$ matrix the boundary
non-local charges follow immediately from the expressions above,
given that the term proportional to $q^{-\theta^-}$ is omitted.
Analogous expressions are obtained for a lower triangular matrix,
and in this case the term proportional to $q^{\theta^-}$ is
omitted in the expressions (\ref{Q}).

It is worth elaborating a bit further on the lattice sine--Gordon.
Our aim is to provide explicit expressions of the lattice
topological charge \cite{izko} and the fractional-spin integrals
of motion for the lattice sine Gordon model. To achieve this we recall
that locally the $L$ matrix (\ref{sg}) lacks a highest (lowest)
weight, and as a consequence the algebraic Bethe ansatz
formulation cannot be applied. However, it was proposed in
\cite{izko} that one may consider a $2N$ site spin chain and
locally can deal with a new $L$ matrix constructed as a
combination of two successive $L$ matrices. Here we are dealing
with a system with open boundaries, and we also need to define the
corresponding $\hat L$ matrix. Indeed we define the following
objects \be {\mathbb L}_{0n}(\lambda) = L_{0\ 2n-1}(\lambda)\
L_{0\ 2n}(\lambda), ~~~\hat {\mathbb L}_{0n}(\lambda) =\hat L_{0\
2n}(\lambda)\ \hat L_{0\ 2n-1}(\lambda) \ee where the subscripts
$2n,\ 2n-1$ in the $L$ operators denote the corresponding site in
the one dimensional lattice. Then by keeping the lowest order
terms in the $\mu \lambda \to \infty$ expansion the ${\mathbb L},\
\hat {\mathbb L}$ matrices reduce to the same form as in
(\ref{d0}). In this case in particular $k_{i},\ e_{i},\ f_{i}$ are
given by the following expressions: \be && k_{1,n} =
e^{-i(\Phi_{2n}- \Phi_{2n-1})}, ~~~~~k_{2,n} = k_{1,n}^{-1} \non\\
&& e_{1,n} =-{im \over
q-q^{-1}}(h_{+}(\Phi_{2n})e^{i\Pi_{2n}}e^{i\Phi_{2n-1}}+h_{-}(\Phi_{2n-1})e^{-i\Pi_{2n-1}}e^{i\Phi_{2n}}),
\non\\ &&  f_{1,n} =-{im \over q-q^{-1}}(
h_{+}(\Phi_{2n-1})e^{i\Pi_{2n-1}}e^{i\Phi_{2n}}+h_{-}(\Phi_{2n})e^{-i\Pi_{2n}}e^{i\Phi_{2n-1}}),
\non\\ && e_{2,n} =-{im \over q-q^{-1}}
(h_{-}(\Phi_{2n})e^{-i\Pi_{2n}}e^{-i\Phi_{2n-1}}+h_{+}(\Phi_{2n-1})e^{i\Pi_{2n-1}}e^{-i\Phi_{2n}}),
\non\\ && f_{2,n} = -{im \over
q-q^{-1}}(h_{+}(\Phi_{2n})e^{i\Pi_{2n}}e^{-i\Phi_{2n-1}}+h_{-}(\Phi_{2n-1})e^{-i\Pi_{2n-1}}e^{-i\Phi_{2n}}).
\label{fc}  \ee In the context of the lattice sine--Gordon the $N$
coproduct (see also Appendix A) $K_{1}^{(N)}$ plays the role of
the lattice topological charge, whereas the coproducts
$E_{1}^{(N)}, \ F_{1}^{(N)}$ play the role of the lattice
fractional-spin integrals of motions, i.e. the lattice analogues
of the classical expressions appearing in \cite{bele}, generating
${\cal U}_{q}(\widehat{sl_{2}})$.
\\
\\
{\bf (II)} {\it The $q$ harmonic oscillator}: The asymptotic
expansion of $L(\lambda -\Theta)$ and $\hat L(\lambda +\Theta)$ as
$\mu \lambda \to \infty$ gives rise to the following matrices \be
&& L_{0n}(\mu \lambda \to \infty) \propto \left(\begin{array}{cc}
V_{n} &\\
      &0 \\ \end{array} \right ) +w'e^{-\mu \lambda} \left(\begin{array}{cc}
 &a_{n}^{+} \\
a_{n}      & \\ \end{array} \right )+ \ldots, \non\\ &&\hat L_{0n}(\mu \lambda \to \infty) \propto \left(\begin{array}{cc}
V_{n} &\\
  &V_{n}^{-1} \\ \end{array} \right ) +\hat w' e^{- \mu \lambda} \left(\begin{array}{cc}
     &a_{n}^{+} \\
a_{n}    & \\ \end{array} \right ) + \ldots \label{la} \ee \be
w'=q^{-{1\over 2}+\Theta'}, ~~~~~\hat w'=q^{-{1\over 2}-\Theta'}.
\ee Finally forming the asymptotics of ${\cal T}$, and bearing in
mind (\ref{la}), (\ref{f0}), we obtain the corresponding boundary
non-local charges for the $q$ harmonic oscillator \be {\cal T}(\mu
\lambda \to \infty) \propto \left(\begin{array}{cc}
 &q^{\theta^-}\\
      & \\ \end{array} \right ) +e^{-\mu \lambda} \left(\begin{array}{cc}
Q_{1}^{(N)} & \\
     & Q_{2}^{(N)}\\ \end{array} \right ) + \ldots \ee and the explicit expressions
of the boundary charges, which are new are given by, \be Q_{1} =
q^{-{1\over 2}+\Theta_{1}} A^{+(N)} V^{(N)} +q^{-{1\over
2}-\Theta_{1}}  V^{(N)}\hat A^{(N)} -{e^{-i\mu \xi^{-}}\over
\kappa^-}(V^{(N)})^2, ~~~~Q_{2}^{(N)}= q^{-{1\over 2}+\Theta_{2}}
A^{(N)} (V^{(N)})^{-1} \label{Q1} \ee where we define \be &&
A^{+(N)} =\bigotimes_{n=1}^{N-1} V_{n}\  a_{N}^+, ~~~~A^{(N)}
=a_{1}\ \bigotimes_{n=2}^{N}V_{n},~~~~V^{(N)}
=\bigotimes_{n=1}^{N}V_{n}\non\\ && \hat A^{(N)} = \sum_{n=1}^{N}
V \otimes \ldots \otimes V \otimes a_{n} \otimes V^{-1} \ldots
\otimes V^{-1}. \label{qqb}  \ee Similarly if one considers an
upper (lower) triangular $K^-$ matrix then the corresponding
boundary charges are obtained by ignoring the terms proportional
to $q^{-\theta^-}$ ($q^{\theta^-}$) exactly as in the previous
case.

It was shown in \cite{doikou1, base} that depending on the choice
of the left boundary either $Q^{(N)}_{1}$ or $Q^{(N)}_{2}$ or a
combination of the two is conserved. For instance for the special
case where the left boundary (principal gradation) is \be
K^{+}(\lambda) = diag(e^{\mu(\lambda +i)},\ e^{-\mu(\lambda +i)})
\label{+} \ee the charge $Q^{(N)}_{1}$ is a conserved quantity
\cite{doikou1}. Actually a stronger statement was shown in
\cite{doikou1}, i.e. the charge $Q_{1}^{(N)}$ turns out to be also
the centralizer of the boundary Temperley--Lieb (blob) algebra in
the spin ${1\over 2}$ XXZ representation. This immediately implies
the commutation of the corresponding Hamiltonian, written in terms
of the blob algebra generators, with $Q_{1}^{(N)}$. The derived
non-local charges will turn out to play a crucial role in the
identification of the spectrum of the corresponding models, hence
they are not only of mathematical but of physical significance as
well. Their relevance to the spectrum will be examined in detail
in the subsequent section.

\section{Diagonalization of the transfer matrix and Bethe ansatz}

\subsection{The pseudo-vacuum}

To identify the spectrum and formulate the Bethe ansatz equations
of the models under consideration we shall apply the method
employed in \cite{chin, FT2}. In the present section we deal with
the XXZ model and the $q$ harmonic oscillator, in the cyclic
representation (\ref{cyclic}), with non-diagonal boundaries. The
lattice sine--Gordon and Liouville models with diagonal boundaries
only will be treated in a subsequent section. Let us give a brief
account of the method, however for a more detailed analysis on the
subject we refer the reader to \cite{chin}. One first introduces
suitable gauge transformations known also as Darboux matrices (see
e.g. \cite{sklyanind})
\begin{equation}
M_{n}(\lambda) = \Big (X_{n}(\lambda), ~~Y_{n}(\lambda) \Big),
~~\bar M_{n}(\lambda) = \Big (X_{n+1}(\lambda), ~~Y_{n-1}(\lambda)
\Big) \label{matrix2} \end{equation} with \begin{equation}
X_{n}(\lambda)= \left (
\begin{array}{c}
  e^{-\mu \lambda}x_{n}\\
  1 \\
\end{array}
\right)\,, ~~ \ \ \ Y_{n}(\lambda)= \left (
\begin{array}{c}
  e^{-\mu \lambda}y_{n}\\
  1 \\
\end{array}
\right)\,, \qquad x_{n}=x_0e^{- i\mu n} \ , \quad y_{n}=y_0e^{i\mu
n}\ .\label{col}
\end{equation}
Here $\{x_0,y_0 \}\in {\mathbb C}$ are $\lambda-$independent and
will be determined explicitly in the following. Introduce also the
matrices \be M_{n}^{-1}(\lambda) = \left ( \begin{array}{c}
\bar Y_{n}(\lambda)\\
\bar X_{n}(\lambda)\\
\end{array}
\right)\,,~~\bar M_{n}^{-1}(\lambda) = \left ( \begin{array}{c}
  \tilde Y_{n-1}(\lambda)\\
  \tilde X_{n+1}(\lambda)\\
\end{array}\right)\ ,
\ee with
\begin{eqnarray} \bar X_{n}(\lambda)= {1\over x_{n} -y_{n}}
\Big (-e^{\mu \lambda}, ~~x_{n} \Big  ), ~~ \bar Y_{n}(\lambda)=
{1\over x_{n} -y_{n}}\Big (e^{\mu \lambda}, ~~-y_{n}\Big  )\ ,
 \non\\ \tilde X_{n}(\lambda)= {1\over x_{n} -y_{n-2}}
\Big (-e^{\mu \lambda}, ~~x_{n} \Big  ), ~~ \tilde Y_{n}(\lambda)=
{1\over x_{n+2} -y_{n}}\Big (e^{\mu \lambda},
 ~~-y_{n}\Big  )\ . \label{col2} \end{eqnarray}
One can check that certain combinations of (\ref{col}),
(\ref{col2}) lead to orthogonal relations among the various vectors. Also, face-vertex
correspondence relations corresponding to commutation relations
between the above quantities can be derived using the $R-$matrix
(\ref{r}), which are omitted here for brevity \cite{chin}. We now
introduce the gauge transformed $L-$operator,
\begin{eqnarray}
{\overline L}_{n}(m|\lambda) &=& \bar M_{(n-1)g+m}^{-1}(\lambda)
L_{n}(\lambda)\bar M_{ng+m}(\lambda)\equiv \left(
        \begin{array}{cc}
 \tilde \alpha_{n} &\tilde \beta_{n} \\
 \tilde  \gamma_{n}  & \tilde  \delta_{n}\\
\end{array}\right)\ ,\non\\ S_{n}(m|\lambda) &=&
M_{ng+m}^{-1}(-\lambda) L^{-1}_{n}(-\lambda)
M_{(n-1)g+m}(-\lambda)\equiv \left(
        \begin{array}{cc}
 \tilde \alpha'_{n} &\tilde \beta'_{n} \\
  \tilde \gamma'_{n}  & \tilde  \delta'_{n}\\
\end{array}\right)\ . \label{gauget}\end{eqnarray}
where $g$ depends on the choice of representation. A priori one
may associate a different representation at each site of the
chain, here however we consider for the sake of simplicity the
case where all sites correspond to the same representation. The
occurrence of a different $g$ at each site, depending on the
choice of representation, is a non trivial observation
facilitating the generalization of the formulation described \cite{chin},
where all sites are associated to the fundamental representation.
Note for instance, that for the spin ${1\over 2}$
representation the factor $g$ is unit, for the spin ${\mathrm s}$
representation is $2{\mathrm s}$, and for the models we are
considering here in the cyclic representation will be specified
later on. Also at the end of the section 6.2 we shall discuss the
special case where all sites correspond to the spin ${1\over 2}$
and the space attached to the boundary is associated to the cyclic
representation (\ref{cyclic}). This scenario in fact corresponds
to the Hamiltonian described in section 4 (\ref{H1}).

It follows that the
transfer matrix (\ref{transfer2}) can be rewritten with the help
of the aforementioned gauge transformations (\ref{gauget}), and
related orthogonal relations \cite{chin} as
\begin{equation}
t(\lambda) = tr_{0} \Big \{ \tilde K_{0}^{+}(\lambda)\ \tilde {\cal T}_{0}(\lambda) \Big \}
\,, \label{gtransfer} \end{equation}
where
\begin{equation}
\tilde K^{+}(m|\lambda) =  \left(
        \begin{array}{cc}
 \tilde K_{1}^{+}(m|\lambda) &\tilde K_{2}^{+}(m|\lambda)  \\
  \tilde K_{3}^{+}(m|\lambda)  &\tilde K_{4}^{+}(m|\lambda)  \\
\end{array}\right)\ = \left(
\begin{array}{cc}
\bar Y_{m}(-\lambda)K^{+}(\lambda) X_{m}(\lambda) &\bar Y_{m}(-\lambda)K^{+}(\lambda)Y_{m-2}(\lambda)  \\
\bar X_{m}(-\lambda)K^{+}(\lambda) X_{m+2}(\lambda)  &\bar X_{m}(-\lambda)K^{+}(\lambda)Y_{m}(\lambda)  \\
\end{array}\right)\ \label{k+}
\end{equation} and \begin{equation}
\tilde {\cal T}(\lambda) =  \left(
        \begin{array}{cc}
  {\cal A}_{m}(\lambda) &{\cal B}_{m}(\lambda)  \\
   {\cal C}_{m}(\lambda)  & {\cal D}_{m}(\lambda)  \\
\end{array}\right)\ = \left(
\begin{array}{cc}
\tilde Y_{m-2}(\lambda) {\cal T}(\lambda)X_{m}(-\lambda) &\tilde Y_{m}(\lambda){\cal T}(\lambda) Y_{m}(-\lambda)  \\
\tilde X_{m}(\lambda){\cal T}(\lambda)X_{m}(-\lambda)  &\tilde X_{m+2}(\lambda){\cal T}(\lambda)Y_{m}(-\lambda)  \\
\end{array}\right)\,.\label{U}
\end{equation}
Similarly to \cite{chin}, one defines the transformed $K^{-}$
matrix as
\begin{equation}
\tilde K^{-}(l|\lambda) =  \left(
        \begin{array}{cc}
 \tilde K_{1}^{-}(l|\lambda) &\tilde K_{2}^{-}(l|\lambda)  \\
  \tilde K_{3}^{-}(l|\lambda)  &\tilde K_{4}^{-}(l|\lambda)  \\
\end{array}\right)\ = \left(
\begin{array}{cc}
 \tilde Y_{l-2}(\lambda)K^{-}(\lambda) X_{l}(-\lambda) &\tilde Y_{l}(\lambda)K^{-}(\lambda)Y_{l}(-\lambda)  \\
\tilde X_{l}(\lambda)K^{-}(\lambda) X_{l}(-\lambda)  &\tilde X_{l+2}(\lambda)K^{-}(\lambda)Y_{l}(-\lambda)  \\
\end{array}\right)\ \label{k-}
\end{equation} with $l=m+Ng$. In the case where a different representation is assigned
at each site then $l= m +\sum_{n=1}^N g_{n}$, where $g_{n}$
characterizes the representation at each site. Note that here the
$K^{\pm}$ matrices are $c$-number solutions of the reflection
equation, with relevant parameters $\xi^{\pm},\ \kappa^{\pm},\
q^{\theta^{\pm}}$.

The first step for the diagonalization of the transfer matrix
(\ref{transfer2}) is the construction of an appropriate
pseudo-vacuum. Such a state will be of the general form: \be
\Omega^{(m)} = \bigotimes_{n=1}^{N} \omega_{n}^{(m)} .
\label{pseudo1} \ee Here we denote $\omega_{n}^{(m)}$ the local
pseudo-vacuum annihilated by the lower left elements of the
transformed matrices (\ref{gauget}), i.e.
\be \tilde \gamma_{n},~\tilde \gamma'_{n}\ \omega_{n}^{(m)}=0. \label{req}\
\label{act}\ee
Such a choice ensures that the operators $T(\lambda)$, $\hat
T(\lambda)$ (\ref{hatmonodromy}), acting on the pseudo-vacuum
state reduce to upper triangular matrices. The above requirements
(\ref{req}) lead to certain sets of algebraic constraints. In
particular, let
 \be L_{0n}(\lambda) =\left(
        \begin{array}{cc}
 \alpha_{n}(\lambda) &\beta_{n}(\lambda)  \\
  \gamma_{n}(\lambda)  &\delta_{n}(\lambda)  \\
\end{array}\right)\ \ee
from the action of $\tilde \gamma_{n}$ on the local pseudo-vacuum
we obtain: \be \Big [-x_{m+ng+1}\ \alpha_{n}(\lambda)+
x_{m+(n-1)g+1}\ \delta_{n}(\lambda) + e^{-\mu \lambda}x_{m+ng+1}\
x_{m+(n-1)g+1}\ \gamma_{n}(\lambda) - e^{\mu
\lambda}\beta_{n}(\lambda) \Big ] \omega^{(m)}_{n} =0 \label{con1}
\ee from the action of $\tilde \gamma'_{n}$ a similar constraint
is entailed.

The non-diagonal elements of (\ref{k+}), (\ref{k-}) acting on the
pseudo-vacuum state (\ref{pseudo1}) must satisfy certain
constraints such that the pseudo-vacuum is indeed an eigenstate of
the transfer matrix (\ref{gtransfer}). Indeed, we associate the
integers $m^{0}$ and $m_{0}$ to the left and right boundaries
respectively, and impose the following:
\be && \tilde K_{2}^{+}(m^{0}|\lambda) =\tilde
K_{3}^{+}(m^{0}|\lambda) = 0\ , ~~~~\tilde K_{3}^{-}(m_{0}|\lambda) = 0\ ,
\label{cond2} \ee
$\tilde K_{2,3}^{\pm}$ are given in (\ref{k+}), (\ref{k-}). The
above constraints may be solved \cite{chin} fixing the relations
among the parameters $m_{0},\ m^0$, $x_0,y_0$. Indeed let $x_{0}
=-ie^{-i\mu (\beta+\gamma )}$, $\ y_{0} =-i e^{-i\mu
(\beta-\gamma)}$, then the relations among the boundary parameters
(\ref{cond2}) are \be && {e^{-i \mu \xi^+} \over 2 \kappa^+} =
-i\cosh i\mu(\beta^+ +\gamma^+ + 1), ~~~{e^{-i \mu \xi^+} \over 2
\kappa^+} = -i \cosh i\mu(\beta^+ -\gamma^+ +1), \non\\ && {e^{-i
\mu \xi^-} \over 2 \kappa^-} = i\cosh i \mu(\beta^- +\gamma^-),
\label{1} \ee and we define \be \gamma^+ = \gamma +m^0,
~~~~\beta^{+} = \beta +\theta^+, ~~~~\gamma^- = \gamma +m_0,
~~~~\beta^{-} = \beta +\theta^-. \label{constraints} \ee For
relevant discussions on these constraints see also \cite{doikouf,
nepo, mart, chin}. It would be illuminating to consider the above
constraints bearing in mind the identifications of the boundary
parameter with the parameters of the blob algebra \cite{doikou1}.
Indeed analogous relations emerge in the double Temperley--Lieb
algebra for exceptional values of the algebra parameters in
\cite{nic}. It is not clear to us however why such constraints
hold true for generic values of the boundary parameters. The
natural question raised is whether this is simply a disadvantage
of the methods employed or it reflects a deeper physical or
algebraic reason. This is an intricate point, which merits further
investigation.

In what follows we examine the pseudo-vacua for both the XXZ
model, and the $q$ harmonic oscillator. Their explicit forms are
specified by solving the formulas (\ref{con1}), which lead to sets
of difference equations.
\\
\\
${\bf (I)}$ {\it The XXZ model}: As already mentioned we shall
treat the cyclic representation (\ref{l1}), (\ref{terms}),
(\ref{cyclic}), for the special case where $q$ is root of unity. We
associate each site of the chain with the cyclic representation,
and we define the local pseudo-vacuum state as
\be \omega_{n}^{(m)} = \sum_{i=1}^{p} w_{l}^{(m, n)}\ f^{(n)}_{l} \label{pseudo} \ee
$f^{(n)}_{l}$ is a $p$ dimensional column vector with zeroes
everywhere apart from the $l^{th}$ position. The constraints
(\ref{con1}) provide (a) the value of $\bf g=2 s-2$ and (b)
recursion relations among the $w_{k}^{(m,n)}$: \be &&
{w_{k+1}^{(m,n)} \over w_{k}^{(m,n)} } = {q^{s-1+\Theta} \sinh
i\mu (k-s+1) \over x_{m+(n-1)g +1}\ \sinh i\mu (k+s)},
~~~~~{w_{p}^{(m,n)} \over w_{1}^{(m,n)} } = x_{m+(n-1)g +1}\ q^{
-\Theta+1 -s} {\sinh (i\mu s) \over \sinh i\mu (1-s)}.
\label{rec3} \ee By normalizing the state with $w_{1}^{(m,n)}=1$
we can write \be w_{k}^{(m,n)} = \Big ({q^{s-1+\Theta}  \over
x_{m+(n-1)g +1}\ } \Big )^{k-1} \prod_{j=1}^{k-1} {\sinh i\mu
(j-s+1) \over \sinh i\mu (j+s)}. \label{rec30}  \ee Comparing the
expressions of $w_{p}^{(m,n)}$ from equations (\ref{rec3}),
(\ref{rec30}) we obtain the following constraint on $x_{m}$, i.e.
\be \Big (-x_{m+(n-1)g+1}\ q^{-\Theta + 1 -s} \Big )^{p} =1. \label{rec300}\ee
Relations (\ref{rec3}) are sufficient for the derivation of the
spectrum.
\\
\\
${\bf (II)}$ {\it The $q$ harmonic oscillator}: For the $L$ matrix
describing the $q$ harmonic oscillator (\ref{l2}) we consider $X
\to q^{z} X,\ Y \to q^{{1\over2}}Y$ (cyclic representation
(\ref{cyclic})). In this case ${\bf g=-1}$. The local
pseudo-vacuum will be also of the form (\ref{pseudo}), and the
corresponding recursion relations are also provided by solving the
constraints (\ref{con1}), \be {w_{k+1}^{(m,n)} \over w_{k}^{(m,n)}
} = - {2 \sinh i\mu(k-z+{1\over 2}) \over x_{m+ng +1}\ q^{-k+z
-1-\Theta}}, ~~~~~{w_{p}^{(m,n)} \over w_{1}^{(m,n)} } = -{
x_{m+ng +1}\ q^{ z -1 -\Theta} \over 2 \sinh i \mu ({1\over
2}-z)}. \label{rec4} \ee Again normalizing the pseudo-vacuum with
$w_{1}^{(m,n)}=1$ we conclude that \be w_{k}^{(m,n)} = \Big (-
{2q^{-z +1+\Theta} \over x_{m+ng +1}}  \Big )^{k-1}
\prod_{j=1}^{k-1} q^{j}\sinh i\mu(j-z+{1\over 2}), \label{rec40}
\ee and the constraint necessary from cyclicity leads to: \be \Big
(-{1 \over 2} x_{m+1+ng}\ q^{z-1-\Theta} \Big )^{p} =
\prod_{k=1}^{p} q^{k} \sinh i\mu (k +{1\over 2} -z). \label{rec400}\ee

To derive the spectrum of the transfer matrix for both the XXZ and
the $q$ harmonic oscillator we also need to know the action of the
transformed diagonal elements on the pseudo-vacuum, which are
given by: \be && \tilde \alpha_{n}^{(m)}\ \omega^{(m)}_{n} =
a(\lambda)\ \omega^{(m+1)}_{n}, \non\\ &&\tilde \delta_{n}^{(m)}\
\omega^{(m)}_{n} =  {x_{m+ng+1} -y_{m+ng-1} \over x_{m+(n-1)g+1}
-y_{m+(n-1)g-1}}\ b(\lambda)\  \omega^{(m-1)}_{n} \non\\ && \tilde
\alpha_{n}^{'(m)}\ \omega^{(m)}_{n} = a'(\lambda)\
\omega^{(m-1)}_{n}, \non\\ && \tilde \delta_{n}^{('m)}\
\omega^{(m)}_{n} =   {x_{m+(n-1)g} -y_{m+(n-1)g} \over x_{m+ng}
-y_{m+ng}}\ b'(\lambda)\ \omega^{(m+1)}_{n}. \label{diag3} \ee The
values of $a$, $\ b$, $\ a'$, $\ b'$ for each representation are
given below \be {\bf (I)} && a(\lambda) = q^{-s+1}\ \sinh \mu
(\lambda +is-i\Theta -i), ~~~~~b(\lambda) =  q^{s-1}\ \sinh \mu
(\lambda -is- i \Theta +i) \non\\ ~~~&& a'(\lambda) = q^{-1+s}\
\sinh \mu (\lambda +is+i\Theta), ~~~~~b'(\lambda) =  q^{1-s}\
\sinh \mu (\lambda -is + i \Theta + 2i) \non\\  {\bf (II)} &&
a(\lambda) = {1\over 2}q^{1-z+\Theta} e^{- \mu \lambda},
~~~~~b(\lambda) =  q^{z-{1\over 2}}\ \sinh \mu (\lambda -i
\Theta+{i\over 2})  \non\\ ~~~&& a'(\lambda) = q^{-{1\over 2}+z}
\sinh \mu (\lambda +{i \over 2}+i \Theta), ~~~~ b'(\lambda) =
{1\over 2} q^{2-z}\ e^{\mu (\lambda+ i \Theta)}. \label{diagg} \ee

\subsection{Spectrum and Bethe ansatz equations}
Now that the appropriate gauge transformations and the
corresponding pseudo-vacua have been explicitly derived, we can
proceed with the computation of the eigenvalues of the transfer
matrix. Namely, we are looking for the solution of the following
eigenvalue problem
\begin{equation}
t(\lambda)\ \Psi = \Lambda(\lambda)\ \Psi
\label{eproblem} \end{equation}
where $\Psi$ is the general Bethe ansatz state of the
form
\begin{equation}
\Psi  = {\cal B}_{m^{0}-2}(\lambda_1) \ldots {\cal
B}_{m^{0} -2M} (\lambda_M)\ \Omega^{(m)} \ ,
\label{bethestate}
\end{equation}
and $m$ will be defined below \cite{sklyanin, chin}. Then one has
to solve a typical eigenvalue problem (\ref{eproblem}) written
explicitly as
\begin{equation}
t(\lambda)\ \Psi = \Big ( K_{1}^{+}(m^0|\lambda) {\cal
A}_{m^0}(\lambda)+ K_{4}^{+}(m^0|\lambda) \tilde {\cal
D}_{m^0}(\lambda)\Big)\ \Psi, \label{eproblem2}
\end{equation}
where \be \tilde {\cal D}_{m}(\lambda) = {\sinh\mu (2\lambda +i)
\sinh \mu (i m +i\gamma +i) \over \sinh (i \mu) \sinh i \mu
(m+\gamma)} {\cal D}_{m}(\lambda) -  {\sinh\mu ( i m
+i\gamma+2\lambda +i) \over \sinh i \mu (m+\gamma)} {\cal A}_{m}
(\lambda) \nonumber \ee and the elements $
K_{1,4}^{+}(m^0|\lambda)$ are given in Appendix B. Using the
definition of ${\cal T}(\lambda)$ in (\ref{transfer2}) and
(\ref{U}), the action of the operators ${\cal A}_{m}(\lambda)$ and
$\tilde {\cal D}_{m}(\lambda)$ on the bulk part of the
pseudo-vacuum (\ref{pseudo1}) may be derived.
Assuming that each site $n$ of the chain is associated to either
the XXZ model (\ref{l1}) or to the $q$ harmonic oscillator
(\ref{l2}) in the cyclic representation (\ref{cyclic}), it is
convenient to define \be {\mathrm f}_{n}(\lambda)=a(\lambda)\
a'(\lambda), ~~~~{\mathrm h}_n(\lambda)=b(\lambda)\
b'(\lambda), \label{fh}\ee $a$, $\ b$, $\ a'$, $\ b'$ are defined in
(\ref{diagg}). Then bearing in mind relations (\ref{diag3}) and
appropriate algebraic relations arising from the reflection
equation (see also \cite{chin}) we obtain:
\begin{eqnarray}
{\cal A}_{m}(\lambda)\ \Omega ^{(m)} &=& \prod_{n=1}^{N}{\mathrm f}_{n}(\lambda)\
K_{1}^{-}(m_{0}|\lambda)\ \Omega^{(m)} \non\\
\tilde {\cal D}_{m}(\lambda)\ \Omega^{(m)} &=&
\prod_{n=1}^{N}{\mathrm h}_{n}(\lambda)\
 K_{4}^{-}(m_{0}|\lambda)\ \Omega^{(m)}, \label{action}
\end{eqnarray}
$m_{0} =m+gN$, and $ K_{1,4}^{-}(m_{0}|\lambda)$ are introduced in Appendix A.

Having in mind the structure of the general Bethe state
(\ref{bethestate}) it is obvious that we need exchange relations
between the operators ${\cal A}_{m}(\lambda)$ and $\tilde {\cal
D}_{m}(\lambda)$ with  ${\cal B}_{l}(\lambda)$, which can be
deduced with the help of the reflection equation (\ref{re}) (see
also \cite{chin}), then one obtains
\begin{eqnarray}
{\cal A}_{m+2}(\lambda_{1}) {\cal B}_{m}(\lambda_{2}) = {\sinh \mu
(\lambda_{1} + \lambda_{2})\ \sinh \mu (\lambda_{1}
-\lambda_{2}-i) \over \sinh \mu (\lambda_{1} -\lambda_{2})\ \sinh
\mu (\lambda_{1} + \lambda_{2}+i)} {\cal B}_{m}(\lambda_{2}) {\cal
A}_{m}(\lambda_{1}) \non\\- {\sinh (2 \mu \lambda_{2})\ \sinh
(i\mu)\ \sinh \mu (\lambda_{1} -\lambda_{2}-im -i\gamma -i) \over
\sinh \mu (im +i\gamma +i)\ \sinh \mu (\lambda_{1} -\lambda_{2})\
\sinh \mu (2\lambda_{2} +i)} {\cal B}_{m}(\lambda_{1}) {\cal
A}_{m}(\lambda_{2}) \non\\ -  {\sinh (i\mu)\ \sinh (i\mu)\ \sinh
\mu (-\lambda_{1} -\lambda_{2}+im +i\gamma ) \over \sinh \mu (im
+i\gamma +i)\ \sinh \mu (\lambda_{1} +\lambda_{2}+i)\ \sinh \mu
(2\lambda_{2} +i)} {\cal B}_{m}(\lambda_{1}) \tilde {\cal
D}_{m}(\lambda_{2}) \label{com1}
\end{eqnarray}
and
\begin{eqnarray}
\tilde {\cal D}_{m+2}(\lambda_{1}) {\cal B}_{m}(\lambda_{2}) =
{\sinh \mu (\lambda_{1} +\lambda_{2}+2i)\ \sinh \mu (\lambda_{1}
-\lambda_{2}+i) \over \sinh \mu (\lambda_{1} -\lambda_{2})\ \sinh
\mu (\lambda_{1} +\lambda_{2}+i)} {\cal B}_{m}(\lambda_{2}) \tilde
{\cal D}_{m}(\lambda_{1}) \non\\- {\sinh \mu (2\lambda_{1}+2i)\
\sinh (i\mu)\ \sinh \mu (\lambda_{1} -\lambda_{2}+im+i\gamma+i)
\over \sinh \mu (im +i\gamma +i)\ \sinh \mu (\lambda_{1}
-\lambda_{2})\ \sinh \mu (2\lambda_{2} +i)} {\cal
B}_{m}(\lambda_{1}) \tilde {\cal D}_{m}(\lambda_{2}) \non\\ +
{\sinh (2\mu \lambda_{2})\ \sinh 2\mu(\lambda_{1} +i) \ \sinh \mu
(\lambda_{1} +\lambda_{2}+im +i\gamma +2i ) \over \sinh \mu (im
+i\gamma +i)\ \sinh \mu (\lambda_{1} +\lambda_{2}+i)\ \sinh \mu
(2\lambda_{2} +i)} {\cal B}_{m}(\lambda_{1}) {\cal
A}_{m}(\lambda_{2})\ . \label{com2}
\end{eqnarray}
In fact all the above exchange relations, arising from the
reflection equation, hold exactly as described in \cite{chin}.
What is only modified is the action of the diagonal elements and
consequently of ${\cal A}_{m}$, $\tilde {\cal D}_{m}$ on the
pseudo-vacuum (\ref{action}), which clearly depends on the choice
of the particular representation. It may be shown that the state
$\Psi$ is indeed an eigenstate of the transfer matrix if we impose
$m\equiv m^0-2M$. Without loosing generality we assume for
simplicity $m^0=0$, and then it immediately follows that \be m_0=
N g-2 M\ .\label{condm1??}\ee
We may now deduce the transfer matrix
eigenvalues by virtue of equations (\ref{eproblem2}),
(\ref{action}), (\ref{com1}), (\ref{com2}) i.e.
\begin{eqnarray}
&&\Lambda(\lambda) =
\Big ( K_{1}^{+}(0|\lambda)K_{1}^{-}(m_{0}|\lambda)  \prod_{n=1}^{N}{\mathrm f}_{n}(\lambda)\ \prod_{j=1}^{M}
{\sinh \mu(\lambda +\lambda_{j})\ \sinh \mu(\lambda
-\lambda_{j}-i) \over \sinh \mu(\lambda +\lambda_{j}+i)\ \sinh
\mu(\lambda -\lambda_{j})} \non\\
&+& K_{4}^{+}(0|\lambda)K_{4}^{-}(m_{0}|\lambda) \prod_{n=1}^{N}{\mathrm h}_{n}(\lambda)\
\prod_{j=1}^{M} {\sinh \mu(\lambda
+\lambda_{j}+2i)\ \sinh \mu(\lambda - \lambda_{j}+i) \over \sinh
\mu(\lambda +\lambda_{j}+i)\ \sinh \mu(\lambda -\lambda_{j})} \Big ) \label{ll}
\end{eqnarray}
provided a certain combination of `unwanted' terms arising from
the commutation relations (\ref{com1}), (\ref{com2}) that appear
in the eigenvalue expression, is vanishing. This is true as long
as $\lambda_{i}$'s satisfy a set of conditions, namely the Bethe
ansatz equations. These equations guarantee also analyticity of
the eigenvalues, and they are written in the familiar form
\begin{equation} H(\lambda_i)\ \prod_{n=1}^{N}{ {\mathrm f}_{n}(\lambda_i-{i\over 2})
\over {\mathrm h}_{n}(\lambda_i-{i\over 2})}  =-\prod_{j=1}^{M} e_{2}(\lambda_i-\lambda_j)\
e_2(\lambda_i+\lambda_j)\ , \label{BAE}
\end{equation} where we define
\be e_{n}(\lambda) ={\sinh\mu (\lambda +{in\over 2}) \over
\sinh\mu (\lambda -{in\over 2})} ~~\mbox{and} ~~H(\lambda) =
{K_{1}^{+}(0|\lambda - {i \over 2})\over  K_{4}^{+}(0|\lambda - {i
\over 2})} { K_{1}^{-}(m_{0}|\lambda - {i \over 2}) \over
K_{4}^{-}(m_{0}|\lambda - {i \over 2})  } \,. \label{hh0}\ee A few comments
are in order at this point. Notice that for the $q$ harmonic
oscillator a factor $q^{-N}$, which reduces to $\pm 1$ for $N =kp$
($k$ integer or half integer), occurs in the left hand side of
the Bethe ansatz equations,  resembling the case of the XXZ chain
with twisted boundary conditions. As realized in \cite{fati} a
similar factor appears also in the Bethe ansatz equations of the
lattice Liouville model with periodic boundary conditions.
Furthermore, for particular values of the parameter $s$ and the
inhomogeneities $\Theta$ i.e. (I) $s=1$, (II) $\Theta=0$ (see also
(\ref{diagg})), the Bethe ansatz equations become degenerate, and
as a consequence the Bethe ansatz formulation is not appropriate
any more for deriving the spectrum. In this case one has to deal
with the relevant Baxter operators and the Baxter equations
\cite{baxter} in order to examine the associated spectra.

The case of dynamical boundaries studied
in a series of papers (see e.g. \cite{sklyanin, ftw},
\cite{zab}--\cite{bajn} and references therein) can be regarded as a special case of the above description.
Indeed, let us consider a typical example of dynamical boundaries,
which is described by the Hamiltonian of (\ref{H1}).
Consider a chain with $N+1$ sites, and assign to all quantum
spaces $n \in \{1, \ldots, N \}$ the spin ${1 \over 2}$
representation, whereas the $N+1$ site is associated to the cyclic
representation, of ${\cal U}_{q}(sl_{2})$ or the $q$ harmonic
oscillator algebra that is we consider a dynamical boundary at the
right end (see also section 4). In fact, the $N+1$ space is the
`dynamical' space $d$ of section 4. Then the spectrum takes the
form (\ref{ll}), but with the functions ${\mathrm f}_n ,\ {\mathrm
h}_{n}$ being (set here for simplicity all the inhomogeneities to
zero): \be && {\mathrm f}_{n}(\lambda-{i\over 2}) =\sinh^{2}\mu
(\lambda +{i\over 2}), ~~~ {\mathrm h}_{n}(\lambda-{i\over 2})=
\sinh^{2} \mu(\lambda -{i\over 2}), ~~~n \in \{ 1, \ldots, N\}
\non\\ && {\mathrm f}_{N+1}(\lambda-{i\over 2}) =a(\lambda-
{i\over 2})\ a'(\lambda-{i\over 2}), ~~~ {\mathrm
h}_{N+1}(\lambda-{i\over 2})= b(\lambda- {i\over 2})\
b'(\lambda-{i\over 2}), \label{sh} \ee and recall $a(\lambda),\ a'(\lambda),\ b(\lambda),\ b'(\lambda)$ are given
in (\ref{diagg}) for both  ${\cal U}_{q}(sl_{2})$ and the $q$
harmonic oscillator. The pseudovacuum for the spin ${1\over 2}$
case have been found in \cite{chin} so we do not repeat them
here. The local pseodovacuum for the site $n$ is given by
(\ref{pseudo}) and (\ref{rec30}) or (\ref{rec40}) depending on the
dynamics at the boundary i.e. ${\cal U}_{q}(sl_{2})$ or $q$
harmonic oscillator respectively. The general Bethe states are
given in (\ref{bethestate}). We could have considered a
dynamical boundary in the left end as well, then ${\mathrm f}_{1}
= {\mathrm f}_{N+1},\ {\mathrm h}_{1} = {\mathrm h}_{N+1}$.

\subsection{Asymptotics and derivation of $M$}

In \cite{nepo, chin} the spin ${1\over 2}$ XXZ model with both
boundaries being non-diagonal is discussed in detail. In this case
via the asymptotic behavior of the transfer matrix it was shown
that the value of the integer $M$ is fixed (see also
\cite{doikouf, mart} for the spin ${\mathrm s}$ representation). Let us now
consider a special left boundary given by (\ref{+}), then the
value of $M$ may be derived in terms of the spectrum of the
conserved quantity $Q^{(N)}_{1}$. Indeed the asymptotic expansion
of the transfer matrix (\ref{ll}) as $\mu \lambda \to \infty$,
taking also into account (\ref{res}), is given for (I) the XXZ
model, and (II) the $q$ harmonic oscillator: \be {\bf (I)}&&
\Lambda (\mu \lambda \to \infty) = -i\ \kappa^-\
e^{2\mu\lambda(N+1)+i\mu (N+1)} \Big ( e^{i\mu( \beta^- +\gamma^-
-2(s-1) N+2M)} +e^{-i\mu( \beta^- +\gamma^- -2(s-1)N+2M)} \Big )
\non\\  {\bf (II)}&& \Lambda (\mu \lambda \to \infty) = -i\
\kappa^-\ e^{2\mu\lambda(N+1)+i\mu (2N+1)} e^{i\mu( \beta^-
+\gamma^- +2M)} \label{con} \ee comparing with the results of section 5, and
taking into account appropriate normalizations for the $K$ and $R$
matrices, we conclude, \be {\bf (I)}&& Q_{1}^{(N)} = i\
\sinh^{-1}(i\mu)\ \Big ( q^{ \beta^- +\gamma^- + 2M
-2(s-1)N}+q^{-\beta^- -\gamma^- -2M +2(s-1)N}\Big ) \non\\ {\bf
(II)} && Q_{1}^{(N)} = 2\ i\ q^{ \beta^- +\gamma^-+2M +N}.
\label{as} \ee The expressions of the boundary charges for both
XXZ and $q$ harmonic oscillator are given by the expressions
(\ref{Q}) and (\ref{qqb}) respectively. Also the sum
$\beta^-+\gamma^-$ is related to the right boundary parameters of
the chain (\ref{1}). From equation (\ref{as}) the upper (lower) bounds of $M$ (integer) are identified by means of the spectrum of
the non-local operator $Q_{1}^{(N)}$, which plays the role of
$q^{s^z}$, the corresponding conserved quantity when diagonal
boundaries are applied in the spin chain. The relation between
$Q_{1}^{(N)}$ and $M$ underlines the significance of the algebraic
object $Q_{1}^{(N)}$ into the derivation of the spectrum of the
open spin chain with special boundary conditions. The
identification of the spectrum of the boundary non-local charge is
an intriguing problem, and some progress has been already achieved
towards this direction, for particular representations in
\cite{nic}, however the problem in its full generality remains
open. In fact, the operator $Q_{1}^{(N)}$ locally consists of
tridiagonal form matrices for the spin ${\mathrm s}$ representation
(\ref{spins}) (we provide an example in Appendix C). The explicit
diagonalization of $Q_{1}^{(N)}$ for generic number of sites $N$
will be discussed elsewhere.

Similar algebraic objects, forming the
corresponding boundary quantum algebra \cite{doikouh}, are
necessary for the identification of the spectrum of models
associated to higher rank algebras with non diagonal right
boundary.

\section{Boundary Lattice sine--Gordon and Liouville models}

In this section we give a flavor on the lattice sine--Gordon and
Liouville theories with diagonal boundaries, that is in the
expressions for the $c$-number $K^{\pm}$ matrices we set $~\kappa^{\pm} =0$.
In particular, we identify the spectrum and the corresponding
Bethe ansatz equations for both models.

In what follows we consider for simplicity the inhomogeneity
$\Theta$ to be zero. We find local pseudo-vacua annihilating the
lower left element of ${\mathbb L},\ \hat {\mathbb L}$, being
effectively `highest weight' states. Let such a state be of the
form \cite{izko, fati} \be \omega_{n} = f(\Phi_{2n})\
\delta(\Phi_{2n} -\Phi _{2n-1} -\mu) \label{pseudo0} \ee \be \gamma_{n},\ \hat
\gamma_{n}\ \omega_{n} =0 \ee and consequently $f(\Phi)$ should be
a solution of the following set of difference equations \be
~\mbox{Lattice sine--Gordon:} &&
f(\Phi+\mu) = -{h_{-}(\Phi)\over h_{+}(\Phi)}f(\Phi) \non\\
~\mbox{Lattice Liouville:} && f(\Phi+\mu)= -h(\Phi) f(\Phi)
\label{pseudo00} \ee which occur exactly as in the bulk case. When
$q$ is root of unity a cyclicity (periodicity) requirement is
necessary to be imposed on the pseudo-vacuum as well, i.e. (see
also \cite{izko, fati}), \be f(\Phi +p\mu) = f(\Phi).
\label{pseudo2}\ee To find the spectrum of the lattice systems we
also need the explicit action of the diagonal entries of the
${\mathbb L},\ \hat {\mathbb L}$ matrices on the local
pseudo-vacuum. Indeed it is straightforward to show that \be  &&
\alpha_{n}(\lambda)\ \omega_{n} = a(\lambda)\ \omega_{n}, ~~~~\hat
\alpha_{n}(\lambda)\ \omega_{n} = a'(\lambda)\ \omega_{n},\non\\
&& \delta_{n}(\lambda)\ \omega_{n} =b(\lambda)\ \omega_{n},
~~~~\hat \delta_{n}(\lambda)\ \omega_{n} =b'(\lambda)\ \omega_{n}
\ee where \be \mbox{Lattice sine--Gordon:} &&  a(\lambda)=
-4m^{2}\cosh \mu (\lambda -{i\over 2} +{ir \over 2})\ \cosh \mu
(\lambda -{i\over 2} -{ir \over 2}) \non\\  &&b(\lambda)=
-4m^{2}\cosh \mu (\lambda +{i\over 2}
+{ir \over 2})\ \cosh \mu (\lambda +{i\over 2} -{ir \over 2}) \non\\
&& a'(\lambda)= -4m^{2}\cosh \mu
(\lambda +{i\over 2} +{ir \over 2})\ \cosh \mu (\lambda +{i\over
2} -{ir \over 2})\
\non\\  && b'(\lambda)= -4m^{2}\cosh \mu (\lambda +{3i\over 2}
+{i r \over 2})\ \cosh \mu (\lambda +{3i\over 2} -{ir \over 2})\   \non\\
\mbox{Lattice Liouville:} && a(\lambda) = -e^{-\mu (\lambda
-{i\over 2})}\ \sinh \mu (\lambda -{i\over 2}) \non\\ &&
b(\lambda) = - e^{-\mu (\lambda +{i\over 2})}\
\sinh \mu (\lambda +{i\over 2}) \non\\
&& a'(\lambda) = e^{\mu (\lambda +{i\over 2})}\ \sinh \mu (\lambda
+{i\over 2}) \non\\ && b'(\lambda) = e^{\mu (\lambda +{3i\over
2})}\ \sinh \mu (\lambda +{3i\over 2}), \label{aab} \ee the
constant $r$ appearing in the sine--Gordon eigenvalues is given by
$2 \cosh i\mu r = (m^{2} +m^{-2})$, it plays the role of
inhomogeneity and it is associated to the mass scale of the
system. The standard algebraic Bethe ansatz may be applied
\cite{sklyanin}, and the spectrum is obtained having the familiar
form of (\ref{ll}) ($K_{1,4}^{-}(m_0|\lambda) \to
K_{1,4}^{-}(\xi^-|\lambda),\ K_{1,4}^{+}(0|\lambda) \to
K_{1,4}^{+}(\xi^+|\lambda)$), while the corresponding Bethe ansazt
equations as customary follow as analyticity conditions on the
spectrum and they have the form (\ref{BAE}), (\ref{hh0}) with \be
&& {\mathrm f}_{n}(\lambda)= a(\lambda)\ a'(\lambda), ~~~~{\mathrm
h}_n(\lambda)=b(\lambda)\ b'(\lambda) \non\\ &&
K^-_{1}(\xi^-|\lambda)\ K_{1}^+(\xi^+|\lambda) = {\sinh \mu (2 \lambda+2i ) \over \sinh \mu(2\lambda +i)}\sinh
\mu(-\lambda +i\xi^-)\ \sinh\mu(\lambda +i\xi^+) \non\\ &&
K^-_{4}(\xi^-|\lambda)\ K_{4}^+(\xi^+|\lambda) = {\sinh (2 \mu \lambda) \over \sinh \mu (2\lambda +i)}\sinh
\mu(\lambda + i\xi^-+i)\ \sinh \mu(-\lambda +i\xi^+ -i),   \label{fh2}
\ee and $a(\lambda),\ a'(\lambda),\ b(\lambda),\ b'(\lambda)$
given by (\ref{aab}) for both the lattice sine-Gordon and
Liouville models. The integer $M$ appearing in the
Bethe ansatz (\ref{BAE}) this time is associated to the spectrum of the
topological charge $K_{1}^{(N)}$ (\ref{copp}), (\ref{fc}) for the
sine-Gordon model. It should be stressed that the identification
of the boundary non-local charges for the case of the open lattice
Liouville theory is an intriguing task, which needs to be further
pursued.

In the same spirit the case where the right boundary is upper
(lower) triangular may be examined. An obvious reference state
still exists, and the Bethe ansatz equations may be easily
extracted being of the familiar form. In this case of course the
corresponding conserved quantity for the lattice sine-Gordon will
be the charge $Q_{1}^{(N)}$ (\ref{Q}) (for a trivial left boundary
(\ref{+})), but with the terms proportional to $q^{-\theta^{-}}$
($q^{\theta^{-}}$) omitted. Finally the more general scenario with
non-diagonal boundaries may be studied employing appropriate local
gauge transformations. The spectrum is anticipated to be of
the form (\ref{ll}), but this time the integer $M$ should be associated
to the corresponding conserved quantity. In general $M$
being associated to the corresponding conserved non-local charge
reflects essentially the underlying symmetry. Moreover, in this
case the derivation of the reference state involves the solution
of more complicated sets of difference equations, and it will be
left for future investigations.

\section{Conclusions}

The
main aim of the present study was the investigation of a
class of open quantum integrable models emerging from the generalized
XXZ model ({\ref{l1}). In this spirit a particular example of a
spin ${1\over 2}$ XXZ Hamiltonian coupled to a dynamical boundary
(\ref{H1}) was first considered. Explicit expressions of the
corresponding dynamical boundary term associated to the two main
algebras i.e the ${\cal U}_{q}(sl_{2})$ and the $q$
harmonic oscillator algebra were derived (\ref{mm1}),
(\ref{mm2}). The main
point is that the expressions of the boundary terms (\ref{mm1}) and
(\ref{mm2}) are generic at this stage that is independent of the
choice of representation, hence they can be exploited for a
variety of dynamical systems (quantum impurities) coupled at the
end if the chain. As a step towards the identification of the
spectrum of the various models under consideration, and after
recalling known expressions of the boundary non-local charges for
the generalized XXZ model, we derived novel expressions of
boundary non-local charges for the lattice sine Gordon model
(\ref{Q}) and (\ref{fc}), and for the $q$ harmonic oscillator
(\ref{Q1}) and (\ref{qqb}). The significance of such quantities
and their relevance to the spectrum is emphasized in section 6.3.

In order to examine the spectra and eigenstates of the XXZ model
and the $q$ harmonic oscillator with non diagonal boundaries in
the cyclic representation (\ref{cyclic}), we generalized the
approach presented in \cite{chin}. The crucial observation is
that although the local gauge transformations (see (\ref{gauget}))
act on the `auxiliary' space ---being always associated to the
fundamental representation of ${\cal U}_{q}(sl_{2})$--- they
depend explicitly on the choice of representation for each quantum
space (site). In particular, there exists a parameter $g$
incorporated in the local transformations (\ref{gauget}), whose
value depends clearly on the choice of representation for the each
site (for the spin ${1\over 2}$ case is unit). Based on this
observation we were able to derive explicit expressions for the
pseudovacua (\ref{pseudo}), (\ref{rec30}), (\ref{rec300}),
(\ref{rec40}), (\ref{rec400}) by solving sets of involved
recursion relations (\ref{rec3}), (\ref{rec4}). In fact both the
value of $g$ and the recursion relations (\ref{rec3}),
(\ref{rec4}) were deduced from the solution of the constraint
(\ref{con1}), which is the fundamental object in this
construction. More precisely, the main requirement imposed was
that the transformed $L$ matrices (\ref{gauget}) become upper
triangular after acting on the suitable local pseudo vacuum, which
eventually led to the constraint (\ref{con1}). Consequently the
Bethe states (\ref{bethestate}) were identified, and the spectra
(\ref{diagg}), (\ref{ll}), (\ref{fh})  and Bethe ansatz equations
(\ref{BAE}) of the aforementioned models were derived. The
derivation of the pseodovacua and Bethe states is of particular
interest especially in the context of computing correlations
functions (see e.g. \cite{maillet}). Moreover in the semiclassical
limit as known Gaudin type Hamiltonians arise, whose eigenstates
are the semiclassical limits of the Bethe states, and they satisfy
Knizhnik-Zamolodchikov equations (see e.g  \cite{bab}). It is a
compelling task to identify the exact type of the
Knizhnik-Zamolodchikov equations for the semiclassical limit of
the models investigated here, this however will be undertaken
elsewhere. We should also point out that within the spirit
described above the Hamiltonian (\ref{H1}), with a fixed
representation (cyclic) at the right end, was treated as a special
case and the corresponding spectrum was presented in (\ref{ll}),
(\ref{sh}).

Another intriguing new result is the link between the spectrum of
the open transfer matrix with a trivial left boundary and a
general non diagonal right boundary, and the conserved quantity
$Q^{(N)}_{1}$ (see (\ref{con}),
(\ref{as})), which plays essentially the role that $S^z$ plays in
models with both diagonal boundaries. The study of its spectrum
was also emphasized and examples on the diagonalization of the one
site charge for the spin ${\mathrm s}$ and the cyclic
representation (\ref{cyclic}) of the ${\cal U}_{q}(sl_{2})$ were
also presented (see Appendix C). Finally, we were able to identify
the spectrum (\ref{ll}) and (\ref{fh2}) and the corresponding
Bethe ansatz equations (\ref{BAE}), (\ref{fh2}) for the lattice versions of
the sine Gordon and Liouville models, but only with diagonal
boundaries. We also realized that  the corresponding pseudovacua
(\ref{pseudo0}), (\ref{pseudo00}), (\ref{pseudo2}) have the same
structure as in the periodic case \cite{izko, fati}. Note that
this is the first time that these models with open boundaries are
examined, so the aforementioned results are quite useful
especially from the point of view of computing the corresponding
exact boundary $S$ matrices and boundary thermodynamic properties.

It is worth emphasizing that all the models under consideration
share a common spectrum form (\ref{ll}). This is somehow
anticipated given that they all belong to the same universality
class, emerging from the generalized XXZ model (\ref{l1}). More
precisely, the terms $\prod_{i=1}^M$ in (\ref{ll}), and
consequently the right hand side of the Bethe asnatz equations
(\ref{BAE}),  are generic that is independent of the choice of
representation. On the other hand the terms of the spectrum
including ${\mathrm f}_{n}, {\mathrm h}_{n}$ and the left hand
side of the Bethe ansatz equations depend clearly on the choice of
representation, hence the various models naturally give rise to
distinct expressions. Indeed notice the various ${\mathrm f}_{n},
{\mathrm h}_{n}$ functions for each model (\ref{diagg}),
(\ref{fh}), (\ref{sh}), (\ref{aab}), (\ref{fh2}). Furthermore, the
spectrum (\ref{ll}), and the left hand side of the Bethe ansatz
equations (\ref{BAE}), depend on the choice of boundaries see e.g.
$K^{\pm}_{1,4}$ in (\ref{ll}), and also the function $H(\lambda)$
(\ref{hh0}), (\ref{fh2}) associated to the corresponding boundary
conditions. It should be finally stressed that in order to obtain
`numbers' for the spectrum one has to solve the corresponding
Bethe ansatz equations. This can be achieved numerically for
finite size chains, or using thermodynamic techniques as $ N \to
\infty$, obtaining consequential (boundary) scattering information
and (boundary) thermodynamic quantities (see e.g. \cite{doikouf}).
This is a particularly interesting aspect, but it is beyond the
intended scope of the present study, and it will be pursued in a
forthcoming work.
\\
\\
\noindent{\bf Acknowledgments:} I am thankful to P. Baseilhac,
R.I. Nepomechie, A. Nichols and V. Rittenberg for illuminating
discussions. I wish also to thank the organizers of the 3rd Annual
Network Meeting, `Integrable Models and Applications' in Santiago
de Compostela, where part of this work was presented. This work is
supported by INFN and the European Network `EUCLID'; `Integrable
models and applications: from strings to condensed matter',
contract number HPRN--CT--2002--00325.

\appendix

\section{The quantum Kac--Moody algebra ${\cal U}_{q}(\widehat{sl_{2}})$}

We briefly review some basic definitions concerning the quantum group structures. Let \be (a_{ij}) =
\left(
\begin{array}{cc}
   2   &-2\\
  -2  &2\\
\end{array} \right)
\,\ee be the Cartan matrix of the affine Lie algebra
${\widehat{sl_{2}}}$ \cite{kac}, and also define \be [x]_{q} =
{q^{x} -q^{-x} \over q-q^{-1}}. \ee Recall that the quantum affine
enveloping algebra ${\cal U}_{q}(\widehat{sl_{2}})\equiv \cal A$
has the Chevalley-Serre generators \cite{jimbo, drinf}  $e_{i}$,
$f_{i}$, $k_{i}$, $i\in \{1,\ 2\}$ obeying the defining relations
\be k_{i}\ k_{j} = k_{j}\ k_{i}, ~~k_{i}\ e_{j}&=&q^{{1\over
2}a_{ij}}e_{j}\ k_{i}, ~~k_{i}\ f_{j}
= q^{-{1\over 2}a_{ij}}f_{j}\ k_{i}, \non\\
\Big [e_{i},\ f_{j}\Big ] &=& \delta_{ij}{k_{i}^{2}-k_{i}^{-2}
\over q-q^{-1}}, ~~i,j \in \{ 1,\ 2 \} \label{1b}\ee and the $q$
deformed Serre relations \be \chi_{i}^{3}\ \chi_{j} -[3]_{q}\
\chi_{i}^{2}\ \chi_{j}\ \chi_{i} +[3]_{q}\ \chi_{i}\ \chi_{j}\
\chi_{i}^{2} -\chi_{j}\ \chi_{i}^{3} =  0, ~~~~\chi_{i} \in
\{e_{i},\ f_{i} \}, ~~~i\neq j. \ee There exists a homomorphism
called the evaluation homomorphism \cite{jimbo} $\pi_{\lambda}:
{\cal U}_{q}(\widehat{sl_{2}}) \to {\cal U}_{q}(sl_{2})$ \be
&&\pi_{\lambda}(e_{1}) = e_{1}, ~~~~~\pi_{\lambda}(f_{1}) =f_{1},
~~~~\pi_{\lambda}(k_{1}) = k_{1} \non\\ && \pi_{\lambda} (e_{2}) =
e^{-2\mu\lambda}c f_{1}, ~~~~~\pi_{\lambda}(f_{2}) =e^{2\mu
\lambda}c^{-1}e_{1}, ~~~~\pi_{\lambda}(k_{2}) = k^{-1}_{1},
\label{eval1} \ee $c$ is a constant.
As mentioned this algebra is also equipped with a coproduct
$\Delta: {\cal A} \to {\cal A} \otimes {\cal A}$, in particular
the generators form the following coproducts \be &&\Delta(y) =
k_{i} \otimes y + y \otimes k_{i}^{-1}, ~~~y\in \{e_{i},\ f_{i} \}
~~~\mbox{and} ~~~\Delta(k_{i}^{\pm 1}) = k_{i}^{\pm 1} \otimes
k_{i}^{\pm 1}. \label{cop} \ee The $L$-fold coproduct is derived
using the recursion relations \be \Delta^{(L)} = (\mbox{id}
\otimes \Delta^{(L-1)})\Delta  \ee and thus one may obtain explicit
expressions of the $L$ coproduct as \be  \Delta^{(L)}(y) =
\sum_{n=1}^{L} k_{i}\otimes \ldots \otimes k_{i} \otimes
\underbrace{y}_{\mbox{ $n$ position}} \otimes k_{i}^{-1} \otimes
\ldots \otimes k_{i}^{-1} ~~~y \in \{e_{i},\ f_{i} \},
~~~~\Delta^{(L)}(k_{i}) =\bigotimes_{n=1}^{L} k_{i,n}.
\label{ncop} \ee

\section{Transformed $K$ matrices}

Explicit expressions for diagonal elements of the transformed $K$
matrices are provided below \be \tilde K_{1}^+(m^{0}|\lambda) =
-{e^{-\mu\lambda} \kappa^{+} \over \sinh (i\mu \gamma^+)} \Big
({e^{i\mu \xi^{+}} \over 2 \kappa^+} \sinh i \mu(1-\gamma^+) +
{e^{-i\mu \xi^{+}} \over 2 \kappa^+} \sinh \mu(2\lambda+i\gamma^+
+i) + 2i \cosh (i\mu\beta^+) \sinh (2\mu \lambda)\Big ) \non\\
\tilde K_{4}^+(m^{0}|\lambda) = {e^{-\mu\lambda} \kappa^{+} \over
\sinh (i\mu \gamma^+)} \Big ({e^{i\mu \xi^{+}} \over 2 \kappa^+}
\sinh i \mu(1+\gamma^+) + {e^{-i\mu \xi^{+}} \over 2 \kappa^+}
\sinh \mu(2\lambda-i\gamma^+ +i) + 2i \cosh (i\mu\beta^+) \sinh
(2\mu \lambda) \Big ) \non\\ \label{r1} \ee \be \tilde
K_{1}^-(m_0|\lambda) &=& {e^{\mu\lambda} \kappa^{-} \over \sinh
i\mu (\gamma^- -1)} \Big ({e^{i\mu \xi^{-}} \over 2 \kappa^-}
\sinh i \mu(-1+\gamma^-) + {e^{-i\mu \xi^{-}} \over 2 \kappa^-}
\sinh \mu(2\lambda-i\gamma^- +i) \non\\ &-& 2i \cosh i\mu(\beta^-
+1) \sinh (2\mu \lambda) \Big ) \non\\ \tilde
K_{4}^-(m_{0}|\lambda) &=& {e^{\mu\lambda} \kappa^{-} \over \sinh
i\mu (\gamma^-+1)} \Big ({e^{i\mu \xi^{-}} \over 2 \kappa^-} \sinh
i \mu(1+\gamma^-) - {e^{-i\mu \xi^{-}} \over 2 \kappa^-} \sinh
\mu(2\lambda+i\gamma^- +i) \non\\ &+& 2i \cosh i\mu(\beta^-+1)
\sinh (2\mu \lambda)\Big ). \label{r2} \ee By restricting our
attention to the case where the left boundary is a trivial
diagonal matrix (\ref{+}), then the transformed diagonal entries
become \be \tilde K_{1}^+(m^{0}|\lambda) = -{e^{-\mu\lambda} \over
2\sinh (i\mu \gamma^+)} \sinh i\mu (1-\gamma^+), ~~~\tilde
K_{4}^+(m^{0}|\lambda) ={e^{-\mu\lambda} \over 2\sinh (i\mu
\gamma^+)} \sinh i\mu (1+\gamma^+). \ee For the derivation of the
spectrum we shall also need certain combinations of the diagonal
transformed elements i.e.,
\begin{eqnarray}
K_{1}^{+}(m|\lambda) &=& \tilde K_{1}^{+}(m|\lambda) + {\sinh\mu
(i\gamma+m+2\lambda +i) \sinh (i\mu) \over  \sinh i \mu
(1+m+\gamma)\sinh\mu (2\lambda +i)}\tilde K_{4}^{+}(m|\lambda)\ , \non\\
K_{4}^{+}(m|\lambda) &=& {\sinh i\mu(m+\gamma) \sinh (i\mu) \over
\sinh i \mu (1+m+\gamma)\sinh\mu (2\lambda +i)}\tilde
K_{4}^{+}(m|\lambda)\ , \label{k1}
\end{eqnarray}
\begin{eqnarray}
K_{1}^{-}(m_{0}|\lambda)
&=& \tilde K_{1}^{-}(m_{0}|\lambda) \non\\
K_{4}^{-}(m_{0}|\lambda) &=& {\sinh i \mu (m_{0}+\gamma+1) \sinh
\mu(2\lambda +i) \over  \sinh i
\mu (m_{0}+\gamma)\sinh (i\mu )}\tilde K_{4}^{-}(m_{0}|\lambda) \non\\
&-& {\sinh \mu (im_{0}+i\gamma+2\lambda +i)  \over  \sinh i \mu
(m_{0}+\gamma)}\tilde K_{1}^{-}(m_{0}|\lambda)\ . \label{k2}
\end{eqnarray}
It is instructive for our purposes here to consider the asymptotic
behavior of $K_{1,4}^{\pm}$. We shall only deal with the case
where $K^{+}$ is a trivial diagonal matrix (\ref{+}). Considering
the case where $\mu \lambda \to \infty$ we conclude that:
\be && K_{1}^{-}(m_{0}|\mu \lambda \to \infty) =-i \kappa^-
e^{3\mu\lambda-i\mu(\beta^- +\gamma^-)}, ~~~~K_{4}^{-}(m_{0}|\mu
\lambda \to \infty) =-{ i \kappa^- \over 2\sinh (i\mu)} e^{5\mu\lambda+i\mu(\beta^- +\gamma^- +2)} \non\\
&&  K_{1}^{+}(m^{0}|\mu \lambda \to \infty) = e^{-\mu \lambda
+i\mu}, ~~~~K_{4}^{+}(m^{0}|\mu \lambda \to \infty) = 2 e^{-3\mu
\lambda -i\mu} \sinh (i\mu). \label{res}  \ee

\section{The boundary charge as a tridiagonal matrix: diagonalization}

In this appendix we consider a particular example of the boundary
operator $Q_{1}^{(N)}$ (\ref{Q}). We deal with its spin ${\mathrm s}$
representation (\ref{spins}) and for only one site. In this case
the operator reduces to an ${\mathrm n} \times {\mathrm n}$
tridiagonal matrix (see also \cite{sklyanin2}) whose
diagonalization will be the main objective in this Appendix
\\
\\
{\bf (A)} Let us first present the boundary operator for one site
$Q_{1}$ (\ref{Q}) in the spin ${\mathrm s}$ representation (\ref{spins}).
Let us set for simplicity \be A_k = q^{-{1\over 2} -\Theta_{1}}
q^{\alpha_{k}} \tilde C_{k} , ~~~~B_k= q^{{1\over 2}
+\Theta_{1}}q^{\alpha_{k}} \tilde C_{k-1}, ~~~~C_{k} = x_{1}\ q^{2
\alpha_{k}}  \ee the boundary operator may be written as \be Q_{1}
=  \sum_{k=1}^{{\mathrm n}-1} A_{k}\ e_{k\ k+1} +
\sum_{k=2}^{{\mathrm n}} B_{k}\ e_{k\ k-1}  + \sum_{k=1}^{{\mathrm
n}} C_{k}\ e_{kk} \ee which is indeed a tridiagonal matrix. Let us
now solve the corresponding eigenvalue problem. Let $\Psi =
\sum_{l=1}^{{\mathrm n}} w_{l}\ f_{l}$ be an eigenstate of $Q_{1}$
then  \be Q_{1}\ \Psi = \epsilon\ \Psi \ee where $\epsilon$ is the
corresponding eigenvalue. The latter equations can be written in a
more explicit form as a tridiagonal (Jacobi) matrix, \be
\left(\begin{array}{ccccccc}
C_{1} &A_{1} &  & & & &\\
B_{2} & C_{2} & A_{2} & & & & \\
  & \ddots &\ddots &\ddots   &  &  &\\
  & & B_{k} & C_{k} & A_{k} & &\\
  & &\ddots  & \ddots  & \ddots  &   &\\
  & & & &  B_{{\mathrm n}-1} & C_{{\mathrm n}-1} & A_{{\mathrm n}-1}\\
  & & & &  &B_{{\mathrm n}} & C_{{\mathrm n}}
\end{array}\right) \left( \begin{array}{c} w_{1}\\ w_{2} \\ \vdots  \\
w_{k} \\ \vdots \\ w_{{\mathrm n}}  \end{array} \right) =  \epsilon
\left( \begin{array}{c} w_{1}\\ w_{2} \\ \vdots  \\ w_{k} \\ \vdots \\
w_{{\mathrm n}}  \end{array} \right).  \label{tr2} \ee Note that
in \cite{sklyanin2} the diagonalization of triadiagonal matrices
and their relation to $q$ hypergeometric series is discussed (also
related to Leonard pairs, Askey-Wilson polynomials etc. see e.g.
\cite{base} and references therein).

One has now to solve the following recursion relations, in order
to determine the factors $w_{l}$ of the eigenstate, \be A_{1}\
w_{2} = \hat C_{1}\ w_{1}, ~~~~~A_k\  w_{k+1}+ B_k\ w_{k-1} = \hat
C_{k}\ w_{k} ~~~k \in \{2, \ldots, {\mathrm n}-1 \},
~~~B_{{\mathrm n}}\ w_{{\mathrm n}-1} =\hat C_{{\mathrm n}}\
w_{{\mathrm n}} \ee $\hat C_{k} = -C_{k} + \epsilon$. Finally
solving the recursion relations we obtain compact expressions
for the factors, i.e. \be && w_{m} = \prod_{j=1}^{m-1} { {\hat
C}_j \over A_{j}} +\sum_{k=1}^{l-1} (-)^k \sum_{j_{k}>j_{k-1}
>....>j_{1}=2,\ j_{k}-j_{k-1}>1}^{m-1}{\hat C_{1} \ldots \hat
C_{j_{1}-2} B_{j_{1}} \hat C_{j_{1}+1} \ldots \hat C_{j_{k}-2}
B_{j_{k}} \hat C_{j_{k}+1} \ldots \hat C_{m-1}  \over A_{1} \ldots
A_{j_{1}-2} A_{j_{1}} A_{j_{1}+1} \ldots A_{j_{k}-2} A_{j_{k}}
A_{j_{k}+1} \ldots A_{m-1}}, \non\\ && m=2l ~~~\mbox{or}
~~~m=2l-1. \ee
\\
{\bf (B)} It is also worth discussing briefly the cyclic
representation. In this case the structure of the boundary
operator becomes more involved and its diagonalization a more
intriguing task. In particular, the boundary operator is now of
the form: \be Q_{1} =  \sum_{k=1}^{p-1} A_{k}\ e_{k\ k+1} + A_{p}\
e_{p1}  + \sum_{k=2}^{p} B_{k}\ e_{k\ k-1}+B_{1}\ e_{1p}
+\sum_{k}^{p}C_{k}\ e_{kk}. \ee Similarly we define: \be A_{k}=
q^{-{1\over 2} -\Theta_{1}-k}\ {q^{-s-k} - q^{s+k} \over q
-q^{-1}}, ~~~~B_{k} = q^{{1\over2} +\Theta_{1}-k}\ {q^{-s+k}
-q^{s-k}\over q -q^{-1}}, ~~~~ C_{k}= x_{1}\ q^{-2k} \ee and the
eigenvalue problem in a matrix form is then written as \be
\left(\begin{array}{ccccccc}
C_{1} &A_{1} &  & & & & B_{1}\\
B_{2} & C_{2} & A_{2} & & & & \\
  & \ddots &\ddots &\ddots   &  &  &\\
  & & B_{k} & C_{k} & A_{k} & &\\
  & &\ddots  & \ddots  & \ddots  &   &\\
  & & & &  B_{n-1} & C_{n-1} & A_{n-1}\\
 A_{p} & & & &  &B_{n} & C_{n}
\end{array}\right) \left( \begin{array}{c} w_{1}\\ w_{2} \\ \vdots  \\ w_{k} \\ \vdots \\
w_{n}  \end{array} \right) =  \epsilon   \left( \begin{array}{c}
w_{1}\\ w_{2} \\ \vdots  \\ w_{k} \\ \vdots \\ w_{n}  \end{array}
\right)  \label{trr} \ee and finally to identify the spectrum and
the corresponding eigenstates one has to solve the subsequent set
of recursion relations: \be A_{k}\ w_{k+1} + B_{k}\ w_{k-1} =\hat
C_{k}\ w_{k}, ~~~~~k\in \{1, \ldots , p \}. \ee

\end{document}